\documentclass[12pt,cite]{article} 
\usepackage{epsfig}
\usepackage{amssymb}
\usepackage{multirow}
\usepackage{txfonts}
\begin{document}
\begin{center}
\noindent { \bf The sensitivity of the next generation of lunar Cherenkov observations to UHE neutrinos and cosmic rays}\\[2em]
C.W. James\footnote{Corresponding author, clancy.james@adelaide.edu.au}, 
R.J. Protheroe\footnote{rprother@physics.adelaide.edu.au}\\
Department of Physics, The University of Adelaide,  Adelaide, SA 5005, Australia
\end{center}

\centerline{\bf Abstract} We present simulation results for the
detection of ultra-high energy (UHE) cosmic ray (CR) and neutrino
interactions in the Moon by radio-telescopes. We simulate the
expected radio signal at Earth from such interactions, expanding
on previous work to include interactions in the sub-regolith
layer for single dish and multiple telescope systems.  For
previous experiments at Parkes, Goldstone, and Kalyazin we
recalculate the sensitivity to an isotropic flux of UHE
neutrinos.  Our predicted sensitivity for future experiments
using the Australia Telescope Compact Array (ATCA) and the
Australian SKA Pathfinder (ASKAP) indicate these instruments will
be able to detect the more optimistic UHE neutrino flux
predictions, while the Square Kilometre Array (SKA) will also be
sensitive to all bar one prediction of a diffuse `cosmogenic', or `GZK',
neutrino flux.

Current uncertainties concerning the structure and roughness of
the lunar surface prevents an accurate calculation of the
sensitivity of the lunar Cherenkov technique for UHE cosmic ray
astronomy at high frequencies.  However, below $200$~MHz we find
that the proposed SKA low-frequency aperture array should be able to
detect events above $56$~EeV at a rate about 30 times that of the
current Pierre Auger Observatory.  This would allow directional
analysis of UHE cosmic rays, and investigation of correlations
with putative cosmic ray source populations, to be conducted with
very high statistics.

~

~\\

\section{Introduction}

The origin of the UHE CR --- protons and possibly atomic nuclei
with observed energies above $10^{18}$~eV and up to at least
$2\times 10^{20}$~eV --- has long remained a mystery.  The
deflection and scattering of CR trajectories in cosmic magnetic
fields makes the flux of all but the highest energy CR appear
isotropic with respect to the Galaxy regardless of their source,
so that measurements of arrival directions cannot reliably be
used for source identification. At the highest energies, the
deflection is less, and this allows the possibility of `seeing'
nearby UHE CR sources.  Arrival directions of UHE CR detected by
the Pierre Auger experiment above $5.6\times 10^{19}$~eV are 
statistically correlated with positions of nearby AGN, which
are in turn representative of the large-scale distribution of
matter in the local universe \cite{AugerScience07}. However, the
flux is extremely low, and so the nature of the sources of UHE CR
within this distribution remains at present unresolved.

An alternative means of exploring the origin of UHE CR is to
search for UHE neutrinos.  As first noted by Greisen
\cite{Greisen} and by Zatsepin \& Kuzmin \cite{Zatsepin_Kuzmin},
cosmic rays of sufficient energy will interact (e.g.\ via pion
photoproduction) with photons of the 2.73~K cosmic microwave
background radiation (CMBR), with the resulting energy-loss
producing a cut-off in the spectrum (the `GZK cut-off') from a
distance source at around $\sim10^{20}$~eV. These same
interactions produce neutrinos from the decay of unstable
secondaries. Several experiments
\cite{Takeda03,Bird95,Connolly06,Abu-Zayyad05} have reported UHE
CR events with energies above $10^{20}$~eV, and therefore a flux
of these `cosmogenic neutrinos', or `GZK neutrinos', is
almost guaranteed.

Significant information on the CR spectrum at the sources is
expected to be preserved in the spectrum of GZK-neutrinos
\cite{Protheroe04} which varies significantly between different
scenarios of UHE CR production. These include acceleration in the
giant radio lobes of AGN, the decay of super-massive dark matter
particles or topological defects, and $Z$-burst scenarios, the
last of which have already been ruled out by limits placed on an
isotropic flux of UHE neutrinos \cite{Gorham04,Barwick06}. Of
course, neutrinos are not deflected by magnetic fields, and so
should point back to where they were produced, with even a single
detection allowing the possibility of identifying the source of
UHE CR. See refs.~\cite{ProtheroeClay2004,
Falcke2004_SKAscienceCase} for recent reviews of UHE CR
production scenarios and radio techniques for high-energy cosmic
ray and neutrino astrophysics. Here we emphasize that in all
models of UHE CR origin, we expect a flux of UHE neutrinos.

UHE neutrino astronomy will be able to provide much needed clues to
the origin of the UHE CR, as would a method to vastly increase
the number of observed CR at UHE. In Section \ref{thelunarcherenkovtechnique}, 
we describe the lunar Cherenkov
technique, which allows both UHE neutrino and CR detection by
observing the Moon using Earth-based radio-telescopes. Section \ref{description_of_modelling} 
details our simulation, which we
use to review observational methods in the light of the next
generation of radio instruments in Section \ref{observationalphenomenology}. 
We calculate the effective apertures of past and future experiments in Section 5, and use these to place limits on
an isotropic flux of UHE neutrinos from past experiments in Section \ref{pastexperiments}.  
For the next generation of telescopes, we
calculate the sensitivity to, and predict neutrino event rates
expected for, various models of an UHE neutrino flux (Section \ref{futureexperiments}), 
and the event rates from the (known)
cosmic ray flux (Section  \ref{cosmicrays}).  
Our results and future
improvements are discussed in Section  \ref{discussion}.

\section{The Lunar Cherenkov Technique}
\label{thelunarcherenkovtechnique}

A high-energy particle interacting in a medium will produce a
cascade of secondary particles, which develops an excess negative
charge by entrainment of electrons from the surrounding material
and positron annihilation in flight.  The charge excess is
roughly proportional to the number of particles in
electromagnetic cascades, which in turn is roughly proportional
to the energy deposited by the cascade.  Askaryan~\cite{Askaryan}
first noted this effect and predicted coherent Cherenkov emission
in dense dielectric media at radio and microwave frequencies
where the wavelength is comparable to the dimensions of the
shower. At wavelengths comparable to the width of the shower, the
coherent emission is in a narrow cone about the Cherenkov angle
$\theta_C=\cos^{-1} (1/n)$ ($n$ the refractive index), while for
wavelengths comparable to the shower length, the coherent emission is
nearly isotropic. This Askaryan effect has now been
experimentally confirmed in a variety of media
\cite{Saltzberg_GorhamSAND01, Saltzberg_GorhamSALT05,
Saltzberg_GorhamICE07}, with measurements of the radiated
spectrum agreeing with theoretical predictions (e.g.\ ref.\
\cite{Munizetal02}). If the interaction medium is transparent to
radio waves, the radiation can readily escape from the medium and
be detected remotely. Since the power in coherent Cherenkov
emission is proportional to the square of the charge excess,
i.e.\ to the square of the energy deposited, extremely high
energy showers should be detectable at very large distances.

The Lunar Cherenkov technique, first proposed by Dagkesamanskii
and Zheleznykh~\cite{Dagkesamanskii}, aims to utilise the outer
layers of the Moon (nominally the regolith, a sandy layer of
ejecta covering the Moon to a depth of $\sim$10~m) as a suitable
medium to observe the Askaryan effect. Since the regolith is
comparatively transparent at radio frequencies, coherent
Cherenkov emission from sufficiently high-energy particle
interactions (specifically, from UHE cosmic ray and neutrino interactions)
in the regolith should be detectable by Earth-based
radio-telescopes. First attempted by Hankins, Ekers \&
O'Sullivan~\cite{Hankins96} using the Parkes radio telescope,
subsequent experiments at Goldstone (GLUE) \cite{Gorham04} and Kalyazin
\cite{Beresnyak05} have placed limits on an isotropic flux of UHE
neutrinos. Observations continue at both Westerbork (WSRT)
\cite{Scholten06} and the Australia Telescope Compact Array
(ATCA; our project), and the technique has been the subject of
several theoretical and Monte Carlo studies
\cite{ZasHalzenStanev92,MunizZas01,GorhamRADHEP01,Beresnyak04}
together with our own recent work \cite{James07}.

Future radio instruments will provide large aperture array (AA)
tile clusters and arrays of small dishes with very broad
bandwidths, with both factors allowing very strong discrimination
against terrestrial radio frequency interference (RFI). The
culmination of the next generation of radio instruments will be
the Square Kilometre Array (SKA), to be completed around 2020,
with smaller pathfinders such as ASKAP (Australian SKA Pathfinder
\cite{ASKAP07}) to be built in the intervening period. In the
meantime, our project aims to perform a series of experiments
with the Australia Telescope Compact Array (ATCA), an array of
six 22~m dishes currently undergoing an upgrade to an eventual
2~GHz bandwidth. Lunar Cherenkov experiments with these
instruments, together with those proposed for LOFAR
\cite{Scholten06}, represent the foreseeable future of the
technique.

\section{Description of Modelling}
\label{description_of_modelling}

Our simulation method is the same as that developed in a previous
paper \cite{James07} to simulate the detection of UHE neutrinos
by the Parkes lunar Cherenkov experiment of Hankins et al.\
\cite{Hankins96}, and except where we explicitly state
otherwise, the simulation methods for both papers are
identical. The simulation uses Monte Carlo methods, generating
UHE particles incident on the Moon at discrete energies, with
lunar impact parameter $r$ sampled with $p(r) \propto r$ for
$0<r<r_m$, the lunar radius. The Moon was approximated to be
spherically symmetric, with the density being constant within zones as given in
Table \ref{table1}, the outer two layers being the sub-regolith (see Section  \ref{subregolith}) 
and the regolith.

We assume equal proportions of
$\nu_e, \nu_{\mu}, \nu_{\tau}$, since we expect oscillations to result in
complete flavour mixing over extragalactic distance scales \cite{Crocker05}.
Neutrinos and anti-neutrinos are treated identically throughout the simulation.
Cross-sections for both charged-current (CC) and neutral-current (NC)
interactions are taken from Gandhi et al.\ \cite{gandhixsections}, with the
interaction inelasticity $y$ (fraction of neutrino energy given to
hadronic showers --- see Section  \ref{modelling_cherenkov_emission})
sampled from the distributions used by Beresnyak \cite{Beresnyak04}. The $\mu/\tau$
generated in $\nu_{\mu}, \nu_{\tau}$ CC interactions are discarded,
since their bremsstrahlung energy-loss rate will be insufficient
to begin detectable cascades. We ignore photo-hadronic interactions
(which allow high-energy bremsstrahlung photons to convert some of the shower
energy into hadronic cascades), so that the $e^{\pm}$ from $\nu_{e}$ CC
interactions are assumed to produce purely electromagnetic showers.
Unless the $e^{\pm}$ emits a bremsstrahlung photon with a significant
fraction of its initial energy, any secondary hadronic showers
from photo-hadronic interactions will be both of low energy and 
strung out along the shower length, so that the whole showers appears
similar to a purely electromagnetic one. Since $\nu_e$ CC interactions
form approximately only two ninths of all primary neutrino interactions,
and most $e^{\pm}$ thereby produced will not emit such a high-energy
photon, this simplification is justified.

Interactions are treated as point-sources of coherent Cherenkov radiation,
with the Cherenkov cone axis being in the direction of the incoming particle.
We parameterise the radiation spectrum as described in Section  \ref{modelling_cherenkov_emission},
with radiation spectra from the coincident hadronic and electromagnetic showers
in $\nu_{e}$ CC interactions calculated separately. We propagate the radiation
using ray-tracing, treating boundaries between media as infinite plane interfaces.
Transmission of the radiation at these boundaries is calculated from the Fresnel
transmission coefficients for each component of the polarisation, and the
solid-angle-stretching factor applicable to radiation from a point source.
We also account for both large-scale surface roughness (Section  \ref{surface_roughness}),
and absorption (Section  \ref{subregolith}).

\subsection{Treatment of Surface Roughness}
\label{surface_roughness}

\begin{figure*}
\centerline{\epsfig{file=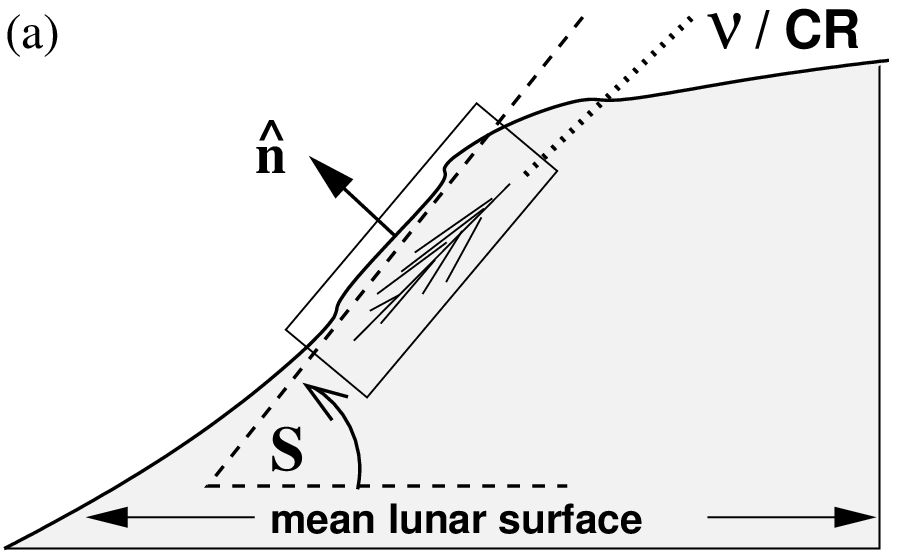, width=7cm} \hspace{1cm}
\epsfig{file=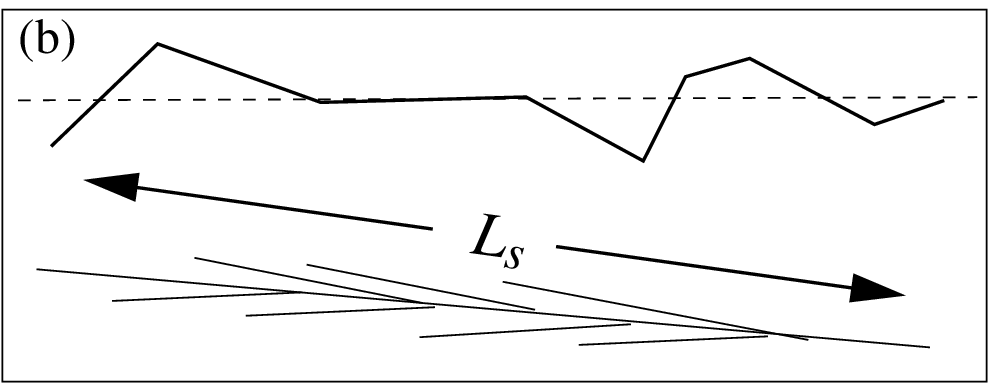, width=7cm}}
\caption{Surface roughness is described by the angular deviation $S$ of the local surface from the
(horizontal) mean lunar surface, assuming a perfectly spherical moon.
(a) Large-scale roughness approximation is illustrated where $S$ is taken to be constant over the length of the shower.
(b) illustrates small-scale roughness where $S$ varies significantly over the length of the shower, $L_s$.}
\label{fig_roughness}
\end{figure*}

The frequency range of the experiments to be modelled is large, and
since the lunar surface is rougher on small length scales than on
large scales the surface will appear smoother at longer
wavelengths. We use the RMS unidirectional slope $S_{\rm rms}$,
i.e.\ the RMS of slopes measured along arbitrary vertical plane
sections through of the surface, based on measured values
appropriate to the observation wavelength of the lunar Cherenkov
experiment in question.  Shepard et al.\ \cite{Shepard95} model
the dependence of $S_{\rm rms}$ on the wavelength as a power-law,
finding the relation between the wavelength $\lambda$ (in cm) and
the slope tangent given in Eqn.\ \ref{rmsslope}:
\begin{eqnarray}
\tan S_{\rm rms} & = & 0.29 \lambda^{-0.22} \label{rmsslope}
\end{eqnarray}
Somewhat confusingly, it is in fact the tangent of $S$ which is
normally distributed (as opposed to $S$ itself), i.e.\ $\tan S$
is sampled from a normal distribution with mean $0$ and standard
deviation $\tan S_{\rm rms}$, though convention dictates that $S$
has the RMS subscript.  In the simulation, the surface normal at
some location is determined as follows: (i) two orthogonal
vertical planes are chosen to intersect with each other and the
nominal spherical lunar surface at the chosen location, (ii) a
line through the intersection point is constructed in each
orthogonal plane, each having a slope sampled from the
distribution above, (iii) the local surface is deemed to be the
surface containing both these lines.  Throughout we use Eqn.\
\ref{rmsslope} with the central frequency $\sqrt{f_1 f_2}$ of
the frequency range $f_1$ to $f_2$ of the observations to
simulate the roughness of the lunar surface. We repeat the
caution of our earlier paper \cite{James07} that this simulates
only the effects of large-scale surface roughness (e.g.\ from
sizeable craters) illustrated in Fig.\ \ref{fig_roughness}(a),
and approximates the effects of small-scale surface roughness
illustrated in Fig.\ \ref{fig_roughness}(b) by using values of
$S_{\rm rms}$ at the scale corresponding to the wavelength of the observations.  In the case of the
Parkes experiment, with a central frequency of $1.4$~GHz and a
bandwidth of 500~MHz, Eqn.\ \ref{rmsslope} predicts $S_{\rm
rms}\approx 8.5^{\circ}$.  In our earlier work we used $S_{\rm
rms}=6^{\circ}$, and given that Beresnyak \cite{Beresnyak04}
found that increasing the surface roughness, i.e.\ increasing
$S_{\rm rms}$, led to a larger effective aperture, the effective
aperture for the Parkes experiment calculated here will thus be
greater than that in ref.~\cite{James07}.

\begin{figure*}
\centerline{\epsfig{file=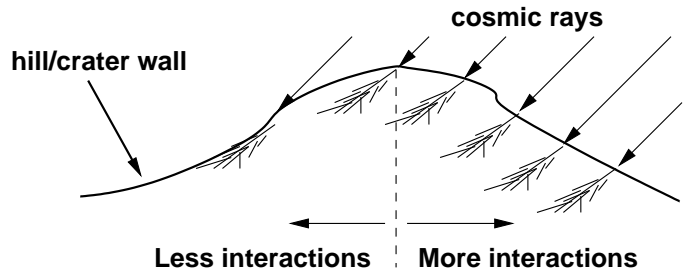, width=9cm}}
\caption{Cosmic rays will tend to interact on the sides of hills where the local surface normal is parallel to the arrival direction, so that the shower develops pointing away from the surface, reducing the probability that radiation from near the Cherenkov angle will escape.}
\label{fig_roughness_crater}
\end{figure*}

To model cosmic rays, the procedure is similar to that for
neutrinos, except that CR are assumed to interact immediately at
the surface, with $100$\% of the energy going into hadronic
showers. Current methods (including ours) of generating $S_{\rm
rms}$ are not really appropriate for cosmic ray interactions,
since correlations between local surface features (which will be
transparent to neutrinos) and the position of cosmic ray
interaction are ignored. As illustrated in Fig.\
\ref{fig_roughness_crater}, cosmic rays will tend to interact
with unfavourable surface slopes, as in the right-hand side of
the hill in the figure, with the surface normal pointing back
towards the particle arrival direction. For such geometries, only
radiation far from the Cherenkov cone will be able to escape
total internal reflection, and the experimental aperture will be
reduced, particularly at high frequencies. One limit (a
`worst-case' limit) of this effect can be made by always
generating the surface normal in the half-hemisphere pointing
back towards the arrival direction of the incident particle,
while a `best-case' estimate comes from ignoring correlations and
generating the surface normal as with neutrinos. We only consider
our estimates of the aperture to UHE CR reliable if the results
of these two methods converge --- where they diverge, a more
thorough treatment will be necessary. As found by Scholten et
al.\ \cite{Scholten05v2}, the suppression of radiation from
showers developing close to the surface (formation zone effect --
see \cite{GorhamRADHEP01}) is expected to be negligible even at
low frequencies where the effect will be greatest, so we ignore
this effect. Note, though, that this suppression is effectively
included in the worst-case limit above, since this method
considerably lowers the probability of generating a shower
developing nearly parallel (and hence close) to the surface.

\subsection{Treatment of the Sub-Regolith Layer}
\label{subregolith}

In our previous simulation,
the medium in which Cherenkov radiation is produced and
transmitted to the lunar surface was modelled only as a single
surface layer, nominally the regolith, with density
$\rho=1.8$~g~cm$^{-3}$. Since the regolith is formed from impact
ejecta, the composition of the regolith --- and hence its
dielectric properties --- should reflect that of the local
underlying material (substrate). As discussed by Wilcox et al.\
\cite{Wilcox06}, what constitutes the boundary between the
regolith and the substrate is poorly defined, since changes in
density (increasing with depth) and the degree of rock
fragmentation (decreasing with depth) are expected to be
smooth. Neither of these should qualitatively affect the
usefulness of the substrate as a dense radio-transparent medium.
Ignoring the sub-regolith layers artificially limits the
aperture, except to low-energy events for experiments observing
at high frequencies, as interactions in the substrate would
otherwise be detectable at low frequencies or high energies. Therefore,
we modify our simulation as described below to allow interactions in a
substrate layer to produce detectable coherent Cherenkov radiation.
Although such a layer has been previously modelled (as an extended,
uniform regolith of 500m depth) \cite{Scholten06}, in our present
work we treat the layers separately, and develop our own model as
described below.

In the Mare, the substrate medium is cooled lava, while in the
highlands, it is the megaregolith. At our current (low) level
of sophistication, a uniform model for both is justified.
Our model, requiring a mean depth and density, dielectric
properties, Cherenkov radiation parameterisation, and a method to
propagate radiation from the substrate to the regolith,
should be more representative of the megaregolith than Mare
basalt, since highland terrain is dominant. We model
the substrate and the $r>1670$~km layer (nominally the crust,
of density $3$~g/cm$^3$ --- see Table \ref{table1}) as
essentially the same medium, which we assume to begin at a depth
of 10~m.  This density is possibly an overestimate, since the
substrate is unlikely to be as dense as the (underlying) crust,
and models for a dual-layer crust (here we use a single layer)
find a lower density for the upper portion \cite{Weiczorek}.  

We use the results of Olhoeft \& Strangway \cite{Olhoeft75} to scale
both the refractive index $n$ and attenuation length $\ell$ of
the substrate from the measured values of the regolith. These
authors obtained the density-dependence of both $n$ and $\ell$ at
frequencies ranging from $100$~kHz to $9$~GHz using results on
soil samples and rock fragments from both Mare and highlands
regions. They also found that the attenuation length depended on
the fraction of iron and titanium present.  While there is
evidence of an increased iron content at depths greater than
$\sim$8~km, no variation at shallower depths has been observed
except from Mare basalts \cite{Lucey95}.  Therefore we scale the
values of $n=1.73$ and $\ell=60~\lambda$ (with $\lambda$ the
wavelength in a vacuum, i.e.\ $\lambda=c/f$) used for the
regolith in our previous paper \cite{James07} solely with the
density as per Olhoeft \& Strangway, obtaining $n=2.5$ and
$\ell=29~\lambda$ for a density of $3$~g/cm$^3$. The depth is
modelled as being essentially infinite, since even at $100$~MHz
the attenuation length of $87$~m is small compared to the
estimated $2$~km mean depth of the substrate \cite{Aggarwal79,
Rasmussen85}.  We believe that our model of the
sub-regolith layer is an adequate reflection of reality, and except
where otherwise noted include such a layer.  Since the nature of
the sub-regolith is poorly constrained by measurement we also
run our simulation with this layer excluded, and it is
feasible that our results for a single shallow layer of regolith will prove to
be the more accurate.

The interface between the sub-regolith and regolith is treated
using ray-tracing as a standard refraction problem at a plane
boundary, on the assumption that the slope of the boundary is the
same as the slope at the surface. This assumption reflects the
likely situation in the Mare, where the lava flows are expected
to present a solid (if fractured) interface.  In the highlands,
the less distinct transition from regolith to sub-regolith will
appear more like the model of a sharp interface between uniform
media at low frequencies, and so will be less sensitive to
inhomogeneities (e.g.\ rock fragments) in the transition
region. At high frequencies, the contribution to the expected
event rate from events below the regolith will be minor due to
absorption in the upper regolith layer, suggesting our
approximation is not inappropriate.  We await the results from
instruments such as the lunar radar sounder \cite{Ono00} on-board
SELENE \cite{Takano05} to enable our model to be improved.

\subsection{Modelling Cherenkov Emission}
\label{modelling_cherenkov_emission}

To calculate the spectrum of observed Cherenkov radiation, we use
the parameterisation developed in our previous paper
\cite{James07}. This combines results from numerous different
papers, in some cases scaled from ice to the regolith and/or from
low to high energies. Since here we also scale the
parameterisation from the regolith to the sub-regolith, in the
interest of clarity, we fully detail all the formula used.

In UHE neutrino-nucleon interactions, the struck nucleon
initiates a hadronic cascade taking typically $\sim$20\% of the
incident neutrino's energy.  In neutral current (NC)
interactions, the remaining $\sim$80\% of the energy goes to the
scattered neutrino which contributes no Cherenkov radiation.
However, in charged current (CC) interactions, a $\mu$, $e$ or
$\tau$ is produced taking the remaining $\sim$80\% of the energy.
A singly-charged $\mu$ or $\tau$ will contribute negligible
Cherenkov radiation, while an electron will initiate an
electromagnetic (EM) cascade.  Thus for neutrinos of energy
$E_{\nu}$ there will be a hadronic shower of energy
$E_S\approx 0.2 E_{\nu}$, and in the case of electron-neutrino CC
interactions there will also be an EM shower of energy
$E_S\approx 0.8 E_{\nu}$. However, such showers will be
substantially lengthened due to the Landau-Pomeranchuk-Migdal
(LPM) effect, with radiation emitted in a correspondingly narrow
cone, so that at very high energies the electromagnetic cascade will actually
contribute less coherent radiation than the hadronic cascade at
most observation angles.  For cosmic rays of energy $E_{CR}$ the
shower energy is simply $E_S=E_{CR}$.

The field strength ${\cal E}_{\rm EM}(\theta_C,f)$ (V/m/MHz) at
distance $R$ (m) from electromagnetic showers 
for radiation emitted at angle $\theta=\theta_C$ from the shower
axis is given by:
\begin{eqnarray}
R \, |{\cal E}_{\rm EM}(\theta_C,f)| & = & V_0 \times \left( \frac{E_S}{1~{\rm EeV}} \right) \, \left( \frac{f}{1~{\rm GHz}} \right) \, \left( \frac{1}{1+ (f/f_0)^{\alpha}} \right) \mbox{~~~~(V/MHz)} 
\label{peak}
\end{eqnarray}
where $f$ is the frequency and $E_S$ is shower energy in EeV
(1~EeV $\equiv 10^{18}$~eV), and $\alpha$ depends on the medium.

The amplitude of coherent Cherenkov radiation from hadronic
showers is related to that of electromagnetic showers by
\begin{eqnarray}
{\cal E}_{H}(\theta_C,f) & = & f(\epsilon) \, {\cal E}_{\rm EM}(\theta_C,f) \label{em_to_h}
\end{eqnarray}
where $f(\epsilon)$ describes the fraction of energy going into
electromagnetic sub-showers, and $\epsilon = \log_{10}(E_S/1~{\rm
GeV})$.  The result for ice \cite{Alvarez-Muniz98} is
\begin{eqnarray}
f(\epsilon) & = & -1.27 \times 10^{-2} - 4.76 \times 10^{-2} \epsilon - 2.07 \times 10^{-3} \epsilon^2 + 0.52\sqrt{\epsilon}. \label{fofeps}
\end{eqnarray}
The function $f(\epsilon)$ is largely determined by the rate of
neutral pion production in hadronic interactions, since neutral
pions decay into electromagnetic products. Variation in
$f(\epsilon)$ is therefore expected between media due to
different hadronic interaction cross sections of the target
nuclei. At very high energies however, the variation is less, and
the dominant source of error comes from uncertainties in the
hadronic interaction models \cite{Alvarez-Muniz04}. Hence Eqn.\
\ref{fofeps} is used unmodified for both the regolith and
sub-regolith.

Away from the Cherenkov angle, the emission falls off due to
decoherence over the width and length of the shower track. This
can be approximated by
\begin{eqnarray}
{\cal E}(\theta,f) & = & {\cal E}(\theta_C, f) \, \left({\sin \theta \over \sin \theta_C}\right) \, 2^{ -(\theta-\theta_C)^2/(\Delta \theta)^2} . 
\label{exponential}
\end{eqnarray}
The $\sin\theta$ dependence becomes significant at low
frequencies, where $\Delta \theta$ is large.  We use $\Delta
\theta$ given by Alvarez-Mu\~{n}iz et al.\ for
electromagnetic~\cite{Alvarez-Muniz97} and
hadronic~\cite{Alvarez-Muniz98} showers, scaling the hadronic
result above $10^{19}$~eV as per Williams \cite{Williams04}. For
hadronic shower energies above $10$~EeV, and for electromagnetic
showers of all energies considered here,
\begin{eqnarray}
\Delta \theta_H & = & C_{\rm H} \left( \frac{f}{1 \; {\rm GHz}} \right)^{-1} \, \frac{1}{1+0.075 \log_{10}(E_S/ 10~{\rm EeV})} \label{hdeltatheta} \\
\Delta \theta_{\rm EM} & = & C_{\rm EM} \left( \frac{f}{1 \; {\rm GHz}} \right)^{-1}\, \left( \frac{E_{\rm LPM}}{0.14E_S+E_{\rm LPM}} \right)^{0.3} 
\label{emdeltatheta}
\end{eqnarray}
where $E_{\rm LPM}$ is the energy above which the LPM effect
becomes important. See Eqn.\ 6 of Alvarez-Mu\~{n}iz et al.\
\cite{Alvarez-Muniz98} for the (rather complex) parameterisation
of $\Delta \theta_H$ for $E_S<10$~EeV showers --- in our
notation, the factor of $1^{\circ}$ in this equation becomes
$0.7^{\circ} \, C_H$, and $\nu_0/\nu$ becomes $1~{\rm GHz}/f$.

The constants $V_0$, $C_H$, $C_{\rm EM}$, $f_0$, $\alpha$ and
$E_{\rm LPM}$ are medium-dependent, with only $E_{LPM}$
calculated explicitly for a given composition \cite{pdg}.
Alvarez-Mu\~{n}iz et al.\ \cite{Alvarez-Muniz06} found the variation
of $\alpha$ between media to be low, and since ${\cal E}$ is
insensitive to changes in $\alpha$ at frequencies below $f_0$,
we adopt the value found by the authors for the regolith
($\alpha=1.23$) also for the sub-regolith. Alvarez-Mu\~{n}iz et al.\
\cite{Alvarez-Muniz06} also used a `box' model of shower
development to predict changes in $V_0$ and parameters related to
$C_{\rm EM}$ and $f_0$ for regolith, given values for ice, and
compared these predictions to the values found from full regolith
simulations.  Variation in all parameters except $V_0$ is
satisfactorily explained by this model, with a $30$\% discrepancy
between the scaled and simulated values of $V_0$.  This is
probably due to the different composition of the regolith when
compared to ice, which would alter the cross-sections at
low-energy for processes in the electromagnetic cascade, the rate
of interactions producing excess track length, and hence the
intensity of coherent Cherenkov radiation. Therefore we expect a
more accurate scaling of $V_0$ from the regolith to sub-regolith
which have identical compositions, and hence changes predicted by
the box model in all electromagnetic shower parameters between
the regolith and substrate are expected to be reliable. Since the
emission from hadronic showers comes from low-energy
electromagnetic cascades, we scale both types of shower
identically.

For the above box-model of shower development, the magnitude of
Cherenkov radiation at the Cherenkov angle is proportional to
$\sin \theta_C / \rho$, where $\rho$ is the density, with both
the width of the Cherenkov cone $\Delta \theta$ and turnover
frequency $f_0$ being proportional to $\rho/\sqrt{n^2-1}$
\cite{Alvarez-Muniz06}.  Using this model, with `R' indicating
regolith and `SR'  sub-regolith, and given $\theta_C^{\rm
SR}=66.4^{\circ}$, we expect $V_0^{\rm SR} = 0.67 V_0^{\rm R}$,
and both $\Delta \theta^{\rm SR} = 1.027 \Delta \theta^{\rm R}$
and $f_0^{\rm SR}=1.027 f_0^{\rm R}$.  Also, the LPM energy
is lower in the sub-regolith due to its higher density.  Table
\ref{table2} details the relevant constants for both layers. The
scaling of these parameters with $\rho$ (since $n=n(\rho)$) means
that a higher sub-regolith density results in a lower effective
aperture for lunar Cherenkov experiments, even once the increased
neutrino interaction rate in a denser medium is accounted
for. The main uncertainty is expected to be due to our limited
knowledge of the structure of the sub-regolith.

\subsection{Modelling Neutrino Detection Experiments}
\label{modelling_experiments}

To model the GLUE and Kalyazin experiments, we use similar
methods to that for Parkes \cite{James07}. We assume Airy beam
patterns, with angular widths calculated from the physical
diameters of the antenna, and estimate the noise in each data
channel / frequency band, assuming a lunar temperature of
$225^{\circ}$~K \cite{troitskij_tikhonova70} and base (or `cold-sky')
system temperature of $35^{\circ}$~K. This method
reproduces well the system temperatures for the past experiments
where reported. For the GLUE experiment, we simulate the full
five-fold coincidence detection algorithm (as implemented for
most of the experiment), with a detection requiring an event to
be above the thresholds specified in Williams \cite{Williams04}
in all five data channels. The simulation for the Kalyazin
experiment is somewhat simpler, needing only to model a
singly-triggered data channel. See Gorham et al.\ \cite{Gorham04}
and Beresnyak et al.\ \cite{Beresnyak05} for details of these
experiments.

It is beyond the scope of this paper to perform a full
optimisation of observation parameters for future experiments
with ATCA, ASKAP, or SKA. To calculate detection thresholds, we
use dual circularly polarised data channels with thresholds of
$V_{\rm thresh} = 6 V_{\rm rms}$ in coincidence, with $V_{\rm
rms}$ calculated as for previous experiments, and cold-sky system
temperatures of $30^{\circ}$ for ATCA \cite{atca_online} and an
assumed value of $35^{\circ}$~K for ASKAP and SKA. Dispersion in
the ionosphere is assumed to be corrected for prior to triggering
(i.e.\ we ignore it), as is beam-forming adequate to cover the
entire Moon with the full collecting area. While this last
assumption is overly optimistic, even for the small number of
baselines provided by ATCA or the expected processing power of
the SKA, the full sensitivity in each case may be recovered by
using a lower trigger threshold off the central core, and writing
buffered data from outlying stations upon triggering.

For reasons outlined below (Section  \ref{observationalphenomenology}), 
future detection
algorithms should search for lunar Cherenkov pulses in multiple
frequency ranges. While for this one pays a statistical penalty
in sensitivity, even a 10-fold increase in statistical trials can
be compensated for by a very small sensitivity loss, e.g.\ by
instead using $V_{\rm thresh}=6.2\, V_{\rm rms}$. For ATCA,
ASKAP, and the SKA, we therefore model a range of frequencies and
bandwidths using $50$~MHz intervals, and choose the experimental
sensitivity to be the highest at a given energy. Bandwidth
limitations are set by the frequency range of the instruments,
except in the case of ASKAP, with a maximum bandwidth of
$300$~MHz. In the case of the SKA, multiple systems will be
utilised to cover the very broad frequency range
\cite{SKAwebsite}. Here we assume three: a sparse aperture array
(AA) at the lowest frequencies (AA low), a dense AA at higher
frequencies (AA high), and dishes for the highest frequencies.
We model the sensitivity separately for each, reflecting the
likely technological constraints of combining the signals from
multiple arrays into a single coherent beam. Table \ref{table3}
 outlines the assumed specifications used to
model the capabilities of these instruments.

\section{Observational Phenomenology}
\label{observationalphenomenology}

The current designs of next-generation radio-instruments are sufficiently
different to that of instruments previously used for lunar
Cherenkov observations to warrant a new investigation of
optimal observational parameters. Here we examine the effects of pointing
position and bandwidth, and explicitly establish
some results which have so far been implicit in the design of
future lunar Cherenkov observations.

\subsection{Broad Bandwidth Observations}
Experiments at Parkes, and in particular Goldstone
and Kalyazin, have operated in the regime where bandwidth $\Delta  f$
is small compared to the observation frequency $f$, so that both
signal field-strength and noise power scale linearly with the bandwidth.
Therefore, the received signal-to-noise power scales linearly with
$\Delta f$, and increasing $f_{\rm max}$ or decreasing $f_{\rm min}$
both equally increase the effective aperture to particles of all energies.

\begin{figure*}
\centerline{\epsfig{file=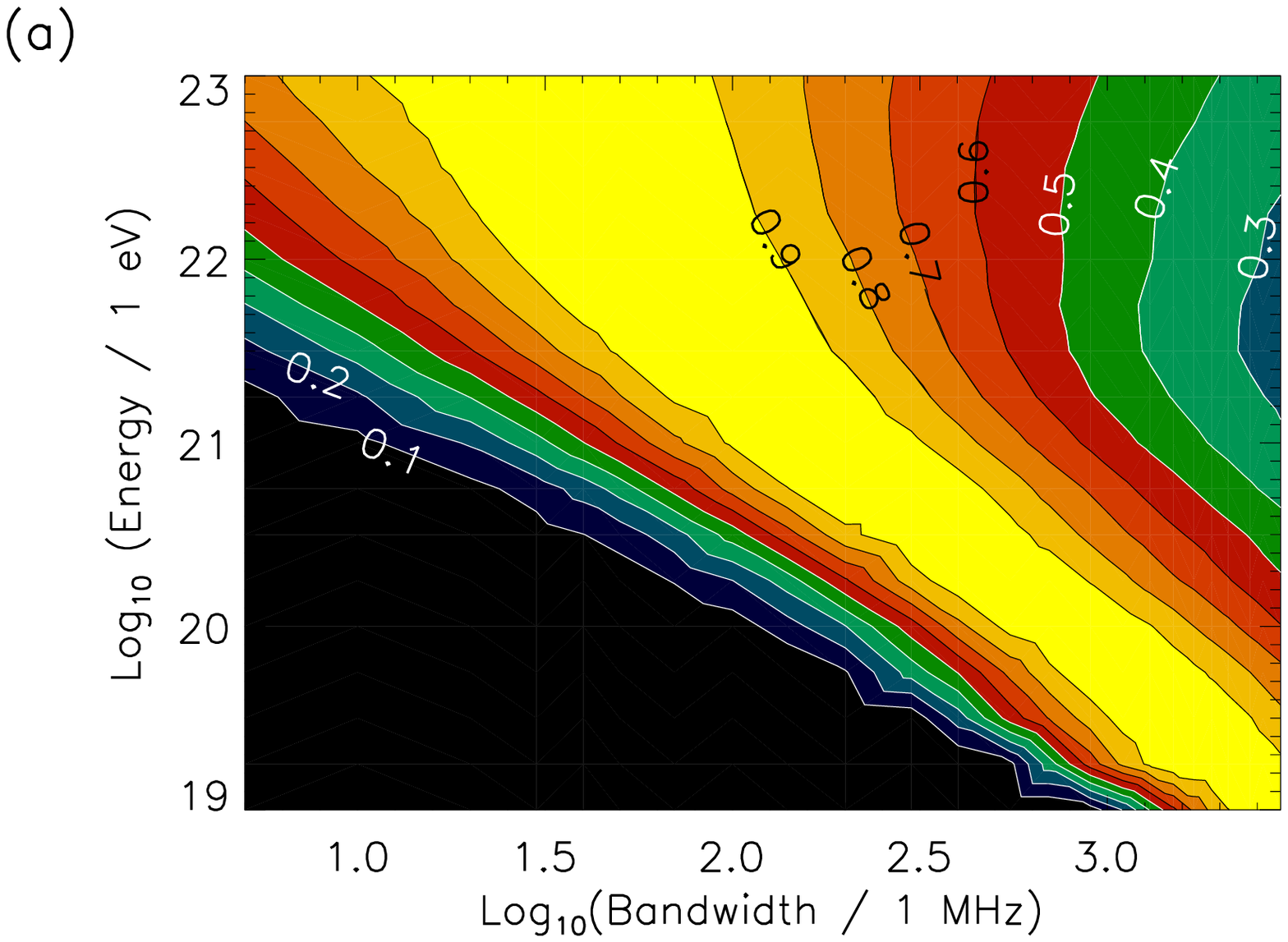, width=8cm} \epsfig{file=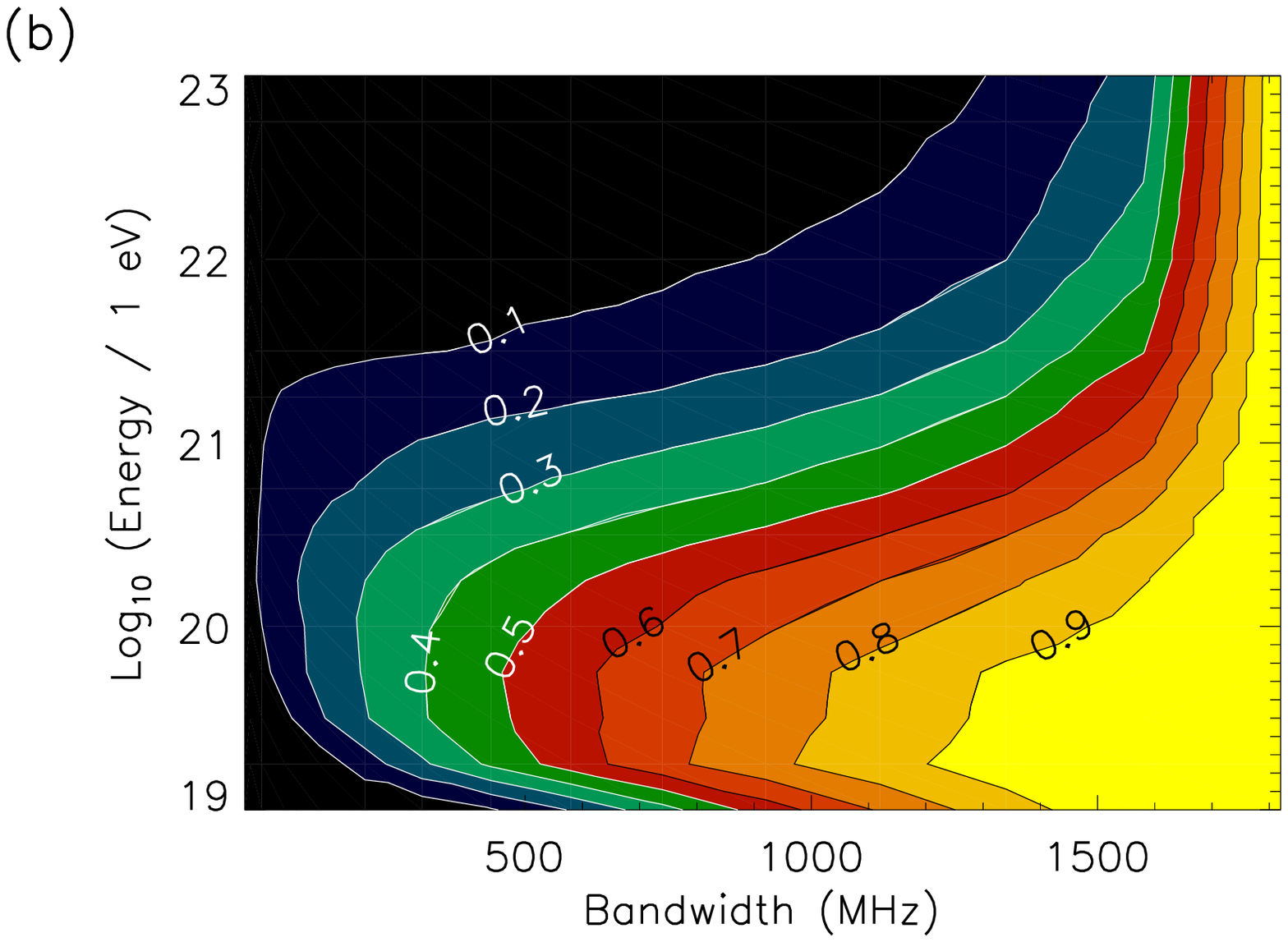, width=8cm}}
\caption{Contour maps of the normalised aperture $A(E_{\nu},\Delta f)/A_{\rm max}(E_{\nu})$ of a fictitious experiment (see text) to UHE neutrinos. (a) for $f_{\rm min}=100$~MHz, (b) for $\bar{f}=1$~GHz.}
\label{bandwidth}
\end{figure*}

This is no longer the case once the bandwidth becomes comparable
to the observation frequency (i.e.\ $\Delta f \nll f$), so that
the signal strength of lunar pulses, and band noise from either
the galactic background (low frequencies) or lunar thermal
emission (high frequencies), may change appreciably over the
bandwidth. In this regime, sensitivity to
geometrically-favourable events (shallow interactions viewed near
the Cherenkov angle) will be increased by extending the bandwidth
to higher frequencies, since coherency from such events could
extend beyond $3$~GHz. However, most geometrically unfavourable
events will be drowned out by the resulting increase in band
noise, since (due to regolith absorption and/or decoherence)
these events will not have a significant high-frequency
signal. To demonstrate this effect, we calculate the effective
area to UHE neutrinos from a fictitious instrument of effective
area $50,000$~m$^2$ and system temperature $50$~K (neglecting
both lunar thermal emission and the galactic background noise),
covering the Moon uniformly in beams. We run simulations over a
variety of bandwidths $\Delta f$, both for fixed $f_{\rm min}$
($=100$~MHz) and fixed mean frequency $\bar{f}=(f_{\rm
max}+f_{\rm min})/2$ ($=1$~GHz). In Fig.\ \ref{bandwidth}, we
show contour plots in the $E_{\nu}$--$\Delta f$ plane of the
aperture $A (E_{\nu}, \Delta f)$ at each energy/bandwidth divided
by the peak aperture for that energy --- (a), for fixed $f_{\rm
min}$; (b), for fixed $\bar{f}$.

The fixed-$f_{\rm min}$ plot (Fig.\ \ref{bandwidth}(a)) clearly
shows that for decreasingly low energy neutrinos, the peak
aperture is achieved with an increasingly high $f_{\rm max}$ and
a broader bandwidth (e.g.\ $\Delta f >1$~GHz for $E_{\nu} <
10^{20}$~eV), since only geometrically-favourable events are
detectable at all. Too low an $f_{\rm max}$ results in an
aperture of zero. At higher energies, the vastly more common
geometrically-unfavourable events become detectable, and the
optimum bandwidth reduces to below $100$~MHz. Importantly, the
decline in aperture by increasing $f_{\rm max}$ above the optimum
is slower than by reducing it below (i.e.\ the contours are
closer together for low bandwidths), even in
log-space. Increasing $f_{\rm max}$ results in a $\sim$linear
increase in noise power, since noise adds incoherently, so the
loss of signal-to-noise through having $f_{\rm max}$ too high is
small.  However, since the signal is coherent, with ${\cal E}(f)$ rising
as steeply as $f$, the rate of loss of peak signal power by decreasing
$f_{\rm max}$ can be up to cubic. Also, at the highest energies,
events appearing only at low frequencies are generally strong
enough to be detected over the increased noise introduced by a
high $f_{\rm max}$.  Thus the reduction in effective aperture for
a lower than optimal $f_{\rm max}$ is worse than for a higher
than optimal $f_{\rm max}$.

The fixed-$\bar{f}$ plot (Fig.\ \ref{bandwidth}(b)) is somewhat
deceptive, and illustrates why in this paper we use $f_{\rm min}, f_{\rm max}$
rather than $\bar{f}, \Delta f$ to describe bandwidths. At first
glace, the figure seems to imply that large bandwidths are ideal at all
energies, which, for a fixed $\bar{f}$, is certainly true. However,
in the high-energy regime, the very strong bandwidth-dependence is due
entirely to the rapid relative variation of $f_{\rm min}$ as $\Delta f
\rightarrow 2 \bar{f}$, because of the aforementioned detection over a
large bandwidth at high energies of strong signals appearing only at
low frequencies. This depends not so much on $f_{\rm max}$ but on
$f_{\rm min}$, which if too high will exclude such events altogether.
However, if the energy is sufficiently low that such events are undetectable,
the effect of increasing bandwidth for fixed $\bar{f}$ is low also, explaining
why the bandwidth-dependence is less at low energies.

\subsection{Optimum Pointing Position}

As first reported by Gorham et al.\ \cite{Gorham01} for GLUE, the
greatest sensitivity to UHE neutrinos for previous experiments
has been achieved when pointing at the lunar limb, partly because
the majority of signals are expected to come from the limb, and
partly because pointing at the limb reduces the lunar thermal
emission received by a detector, thus also reducing the detection
threshold. Both effects become important when the beam size is
comparable to or smaller than the Moon's angular
diameter of $\sim 0.5^{\circ}$. Obviously, if the beam illuminates
the Moon's surface uniformly, the origin of the signals on the lunar surface
does not affect the probability of detection. This is also the case in experiments which can form
multiple independent beams, e.g.\ by the use of focal plane arrays (FPAs).

To model the effects of changing beam size and pointing position, we
simulate the same fictitious instrument as above, but vary the
antenna diameter from $8$ to $256$~m while keeping the total effective collecting
area of the instrument constant (i.e.\ we vary the number of antennas), and add the contribution
of lunar thermal emission to the base system temperature. We (somewhat
arbitrarily) use an observation frequency of $1$~GHz, with a narrow
$100$~MHz bandwidth to remove the aforementioned broad-bandwidth effects.

\begin{figure*}
\centerline{\epsfig{file=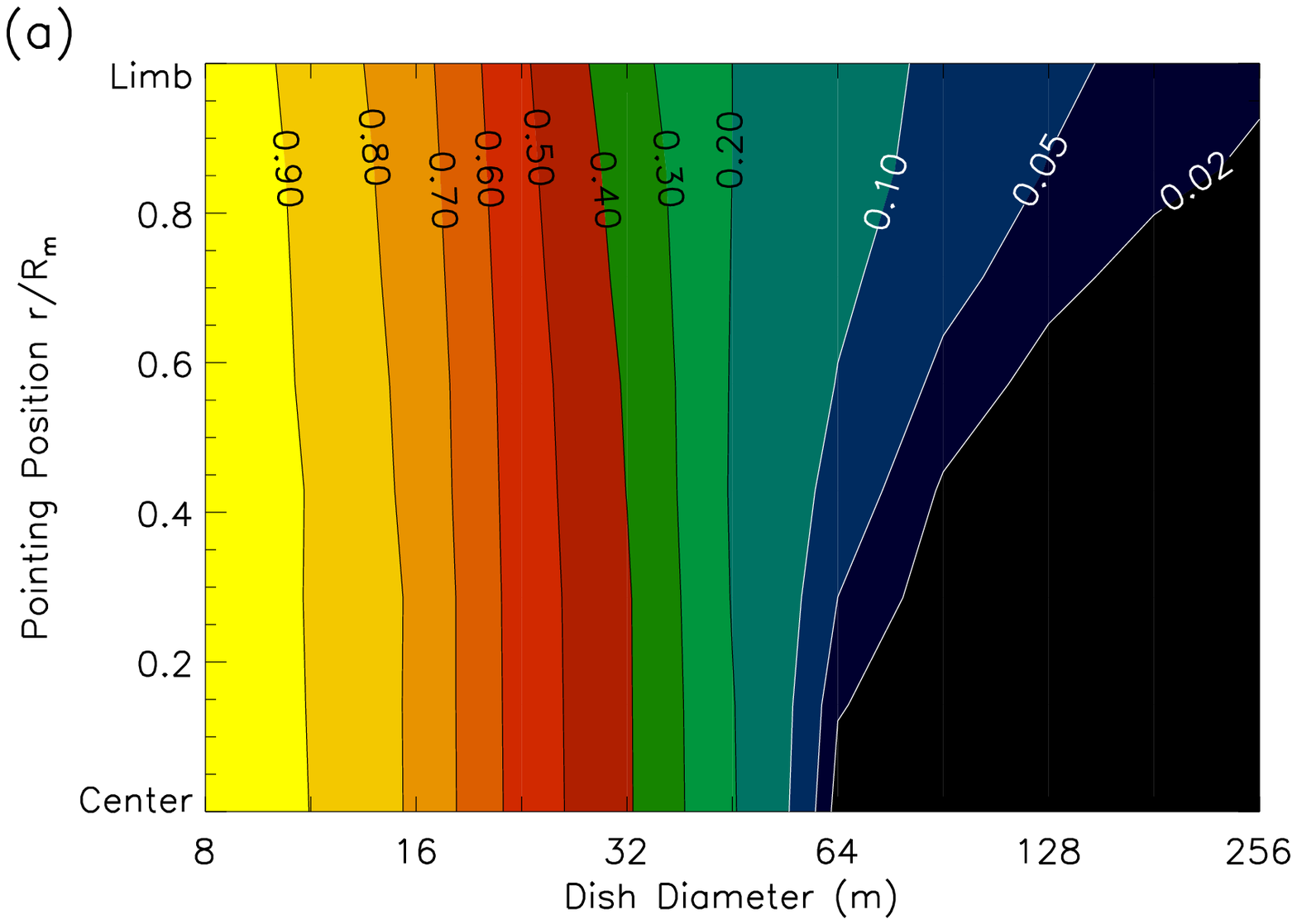, width=8cm} \epsfig{file=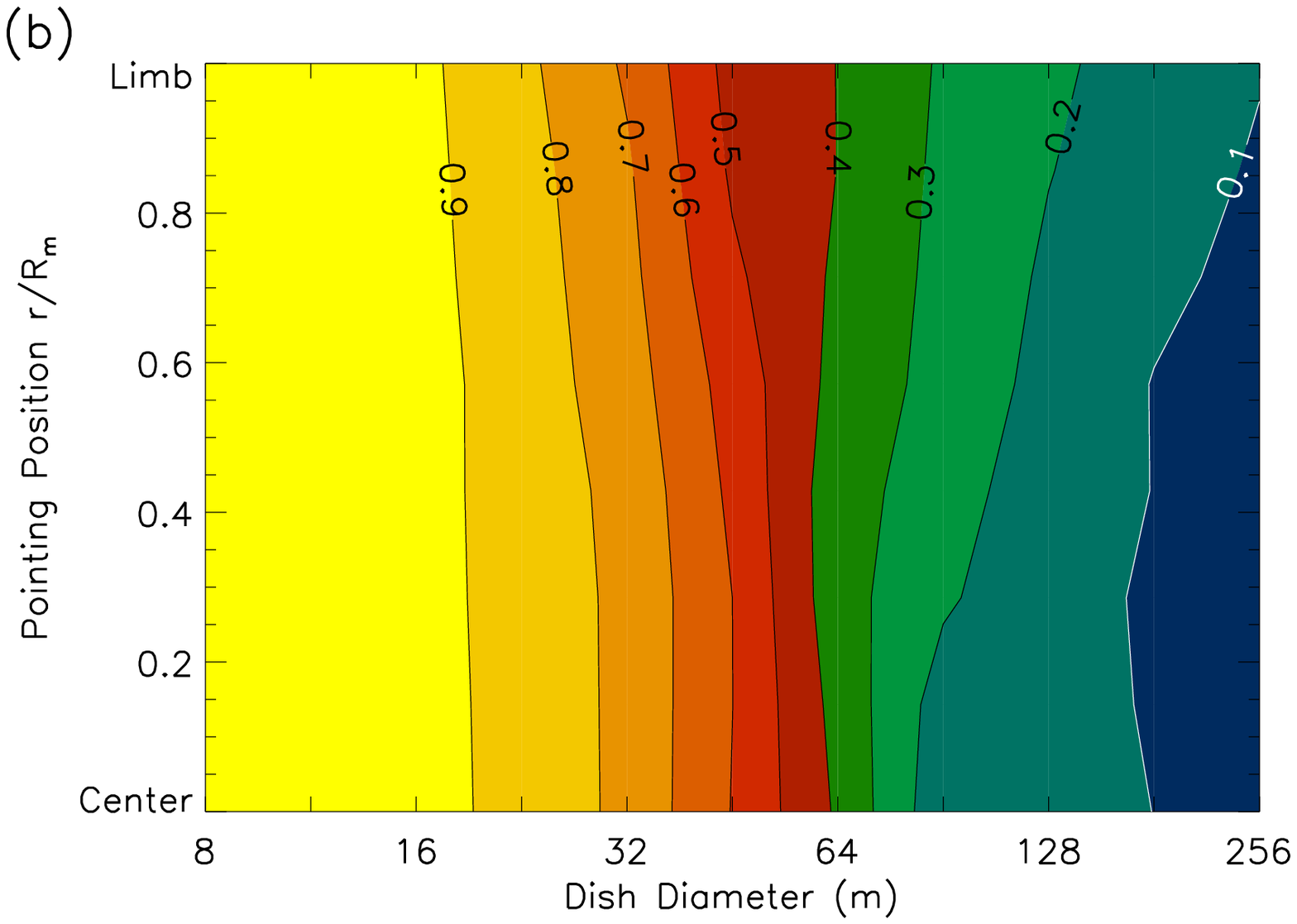, width=8cm}}
\caption{Contour maps of the effective aperture of a fictitious experiment (see text) to UHE neutrinos at energies of (a) $10^{20}$~eV and (b) $10^{22}$~eV, as a function of individual antenna diameter and pointing position. Contour levels are labelled as fractions of the peak aperture, achieved in each case for the centre-pointing mode of the minimum antenna diameter of $8$~m.}
\label{pointings}
\end{figure*}

Our results for $E_{\nu}=10^{20}$ eV and $E_{\nu}=10^{22}$ eV are
plotted in Fig.\ \ref{pointings}. For all energies and pointing
positions, using a larger number of smaller dishes increases the
effective aperture. Smaller dishes have greater coverage of the
lunar surface, due to a larger individual beam size, \textit{and}
an increased sensitivity, since the lunar thermal emission is
largely incoherent between individual dishes. This is one of the
main reasons why ATCA ($6 \times 22$~m dishes) is a potentially
superior instrument over Parkes ($1 \times 64$~m dish), despite a
similar collecting area. Note that in a real experiment, lunar
thermal emission would be partially correlated between individual
antenna. This would increase thermal noise beyond that
calculated, especially for closely-packed small antennas, thus
decreasing their advantage. Also, the processing requirements of
beamforming over the entire lunar solid angle for many individual
antennas are large, which is of course why instruments with a
large number of individual small elements (the `large $N$, small $d$'
concept) are only now being developed.

For large antenna at low energies, pointing at the limb results in
a net gain in aperture, both because the slight reduction in threshold
from reduced lunar noise results in a large increase in detectable
events, and because events far from beam centre (i.e.\ events on the
limb when in centre-pointing mode) are in any case undetectable. At
high energies, the number of events lost from the far limb is greater
than that gained from the near, while the change in sensitivity matters
little since events are energetic enough to be readily detectable.

With these qualitative conclusions established, we now proceed to
calculate the effective apertures for both past and future experiments.

\section{Effective Aperture of Past and Future Experiments}
\label{EffectiveAperture}

\begin{figure*}
\centerline{\epsfig{file=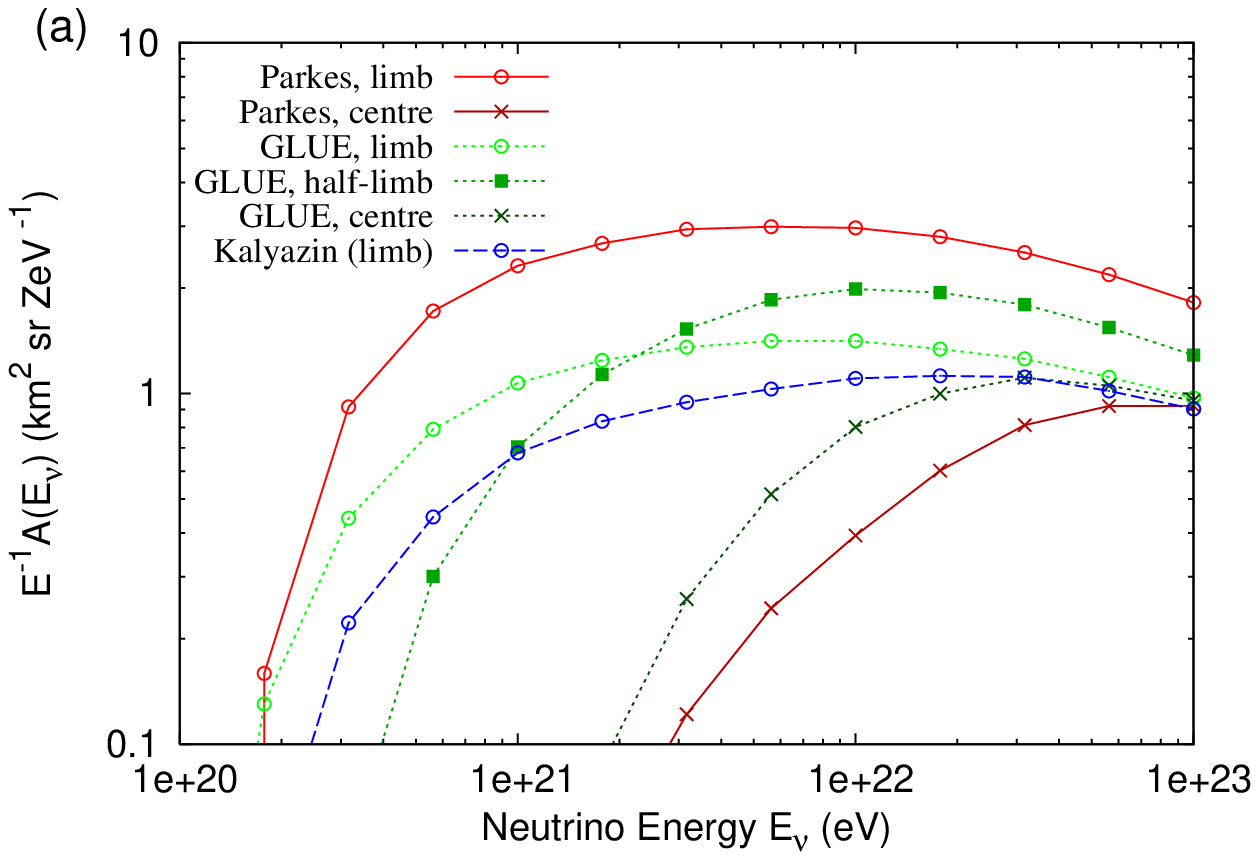,totalheight=5.3cm} \hspace{0.1cm} \epsfig{file=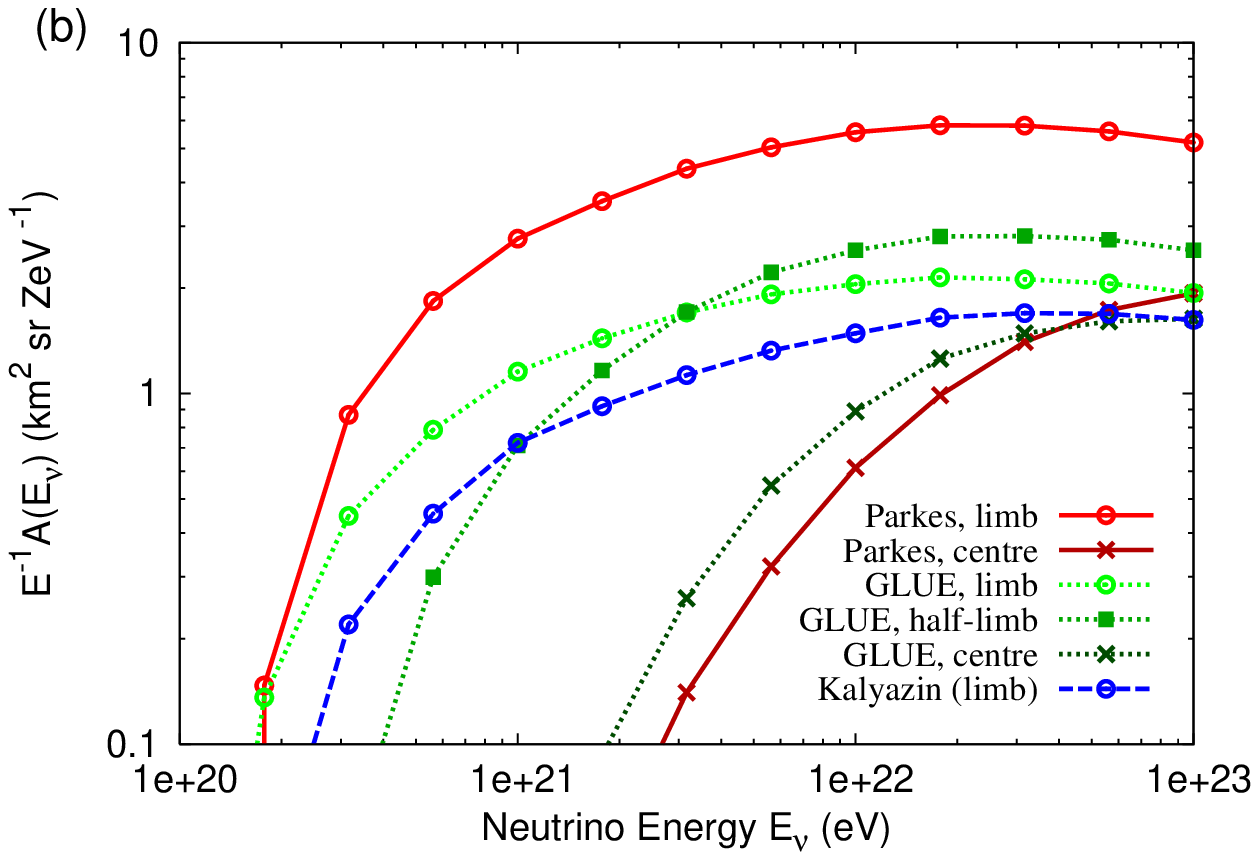, totalheight=5.3cm}}
\caption{Effective apertures to UHE neutrinos of past lunar Cherenkov experiments, divided by neutrino energy in ZeV (1 ZeV $=$ $10^{21}$~eV), both (a) excluding and (b) including the sub-regolith layer.}
\label{past_apps}
\end{figure*}

In Fig.~\ref{past_apps} we plot the effective aperture,
$A(E_{\nu})$ (km$^2$ sr), to neutrinos of previous experiments,
both with (Fig.~\ref{past_apps}(a)) and without (Fig.~\ref{past_apps}(b)) the
sub-regolith layer. For these $>$GHz observations, the
contribution to the effective aperture from interactions in the
sub-regolith dominates only above $\sim 10^{22}$~eV, and so the
dependence of our revised limits on the nature of the
sub-regolith is only significant at energies where some models
predicting large fluxes have already been ruled out
\cite{Barwick06}. As we noted earlier \cite{James07}, the
effective aperture we calculate for GLUE with the sub-regolith
excluded is significantly less than that estimated under similar
assumptions by Gorham et al.\ \cite{Gorham04}.  For reasons we
have noted earlier \cite{James07}, the GLUE and Kalyazin simulations
appear inconsistent with each other, and our result for the
effective aperture of the Kalyazin experiment is in good
agreement with that calculated by Beresnyak et al.\
\cite{Beresnyak04}, after allowing for the greater regolith depth
($30$~m) and lower threshold used.

\begin{figure*}
\centerline{\epsfig{file=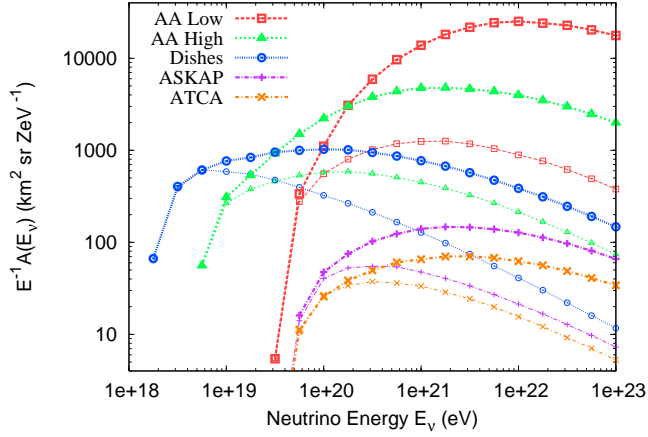,totalheight=6cm}}
\caption{Effective apertures to UHE neutrinos of future lunar Cherenkov experiments, divided by neutrino energy in ZeV (1 ZeV $=$ $10^{21}$~eV), both including (upper, thicker curves) and excluding (lower, fainter curves) the sub-regolith layer.}
\label{future_apps}
\end{figure*}

The predicted effective apertures $A(E_{\nu})$ for future
experiments is shown in Fig.\ \ref{future_apps}. Except in the
case of ASKAP (with a comparatively small bandwidth of $\Delta
f \approx 0.3 f$), the greatest apertures at the highest
energies were indeed achieved by using only a fraction of the
available bandwidth, e.g.\ for the SKA dishes, a full bandwidth
of $0.7$-$3.0$ GHz was optimum for neutrino energies at and below
$2 \times 10^{18}$~eV only, while above $10^{21}$~eV, the
greatest aperture is achieved over a $200$~MHz bandwidth (from
$0.7$-$0.9$~GHz). In all cases, the dependence of the optimal
bandwidth for a given experiment on the inclusion or otherwise of
the sub-regolith layer was very slight, even while the dependence
of $A(E_{\nu})$ was very strong.

Given the large increase in sensitivity offered by the SKA, the
order-of-magnitude reduction in the neutrino-energy detection
threshold over possible experiments with ATCA and ASKAP appears
low, though these in turn offer an order-of-magnitude threshold
reduction over the experiments at Parkes, Glue, and
Kalyazin. Reducing this threshold in an experiment utilising
coherent Cherenkov radiation is so difficult because the emitted
power scales as the square of the shower energy, and the
interaction rate as $E_{\nu}^{0.363}$. The low interaction rate
however means that at $10^{19}$~eV, an aperture of $1$~km$^2$ sr
is equivalent to an effective volume of approximately $40$~km$^3$
water-equivalent seeing all $4 \pi$ sr. We therefore assess the
utility of these experiments in the context of expected event
rates from and potential limits on an UHE neutrino flux.

\section{Neutrino Limits from Past Experiments}
\label{pastexperiments}

\begin{figure*}
\centerline{\epsfig{file=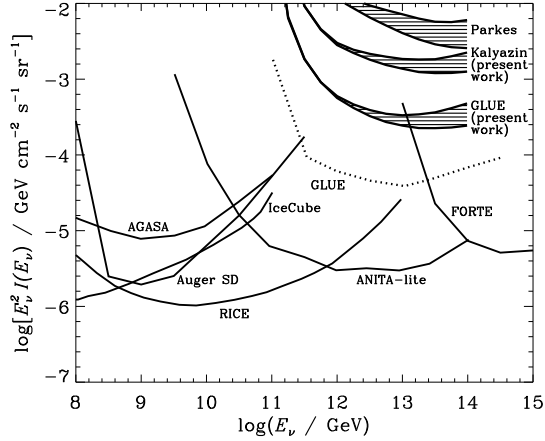, width=90mm}}
\caption{Existing model-independent limits on a total UHE
neutrino flux (adjusted for all neutrino flavours) (see text)
from: GLUE \cite{Gorham04}; IceCube \cite{IceCube07}; RICE
\cite{Kravchenko06}; ANITA-lite \cite{Barwick06}; FORTE
\cite{Lehtinen04}; our revised estimates for Parkes, GLUE, and
Kalyazin are shown by hatched bands (upper boundary -- limit for
10~m regolith; lower boundary -- 10~m regolith plus 2~km
sub-regolith); Auger surface detectors \cite{auger_nu}.}
\label{isotropic_limits}
\end{figure*}

Limits from past experiments have either been expressed as a
`model-independent' limit as per \cite{Lehtinen04} or as a limit
on a (typically) $dN/dE \propto E^{-2}$ flux between two energies
--- we favour the former method, since the resulting limit
reflects the energy-dependence of the effective experimental
aperture. In the case of a non-detection, the corresponding limit is
$s_{\rm up} / [t_{\rm obs} A(E_{\nu})]$, with $t_{\rm obs}$ the
observation time, and the statistical factor for an upper bound
$s_{\rm up}$ at $90$\% confidence is $2.3$ (in the sole case of
FORTE, $s_{\rm up}=3.89$ for one uncertain event). Existing
limits, summed over neutrino flavour, on an isotropic,
uniform-flavour flux of UHE neutrinos are plotted in Fig.\
\ref{isotropic_limits}.  Over the $10^{8}$--$10^{15}$~GeV energy
range, the strongest limits come from IceCube \cite{IceCube07},
RICE \cite{Kravchenko06}, ANITA-lite \cite{Barwick06} and FORTE
\cite{Lehtinen04}.  Our revised estimates for the limits from
GLUE, Kalyazin, and Parkes are now of mostly historical
significance, with our limit for GLUE being approximately an
order of magnitude higher (i.e.\ less limiting) than given in
\cite{Gorham04}, reflecting our lower estimate of the GLUE
effective aperture.

For the projected ANITA limits \cite{Barwick06}, we scale the
published $50$-day sensitivity estimates to the 18 day duration
of their December 2006 balloon flight, and convert to a 90\%
confidence model-independent limit by multiplying by 2.3 because
of presumed non-detection -- we await with interest the
publication of the 2006 result.

\section{Flux Predictions and Future Experiments}
\label{futureexperiments}

We shall first consider the diffuse UHE neutrino intensity
predicted in various models, and then the sensitivity and
expected event rate for future experiments.

\subsection{UHE neutrino predictions and expected event rates}

\begin{figure*}
\centerline{~~~~~~\epsfig{file=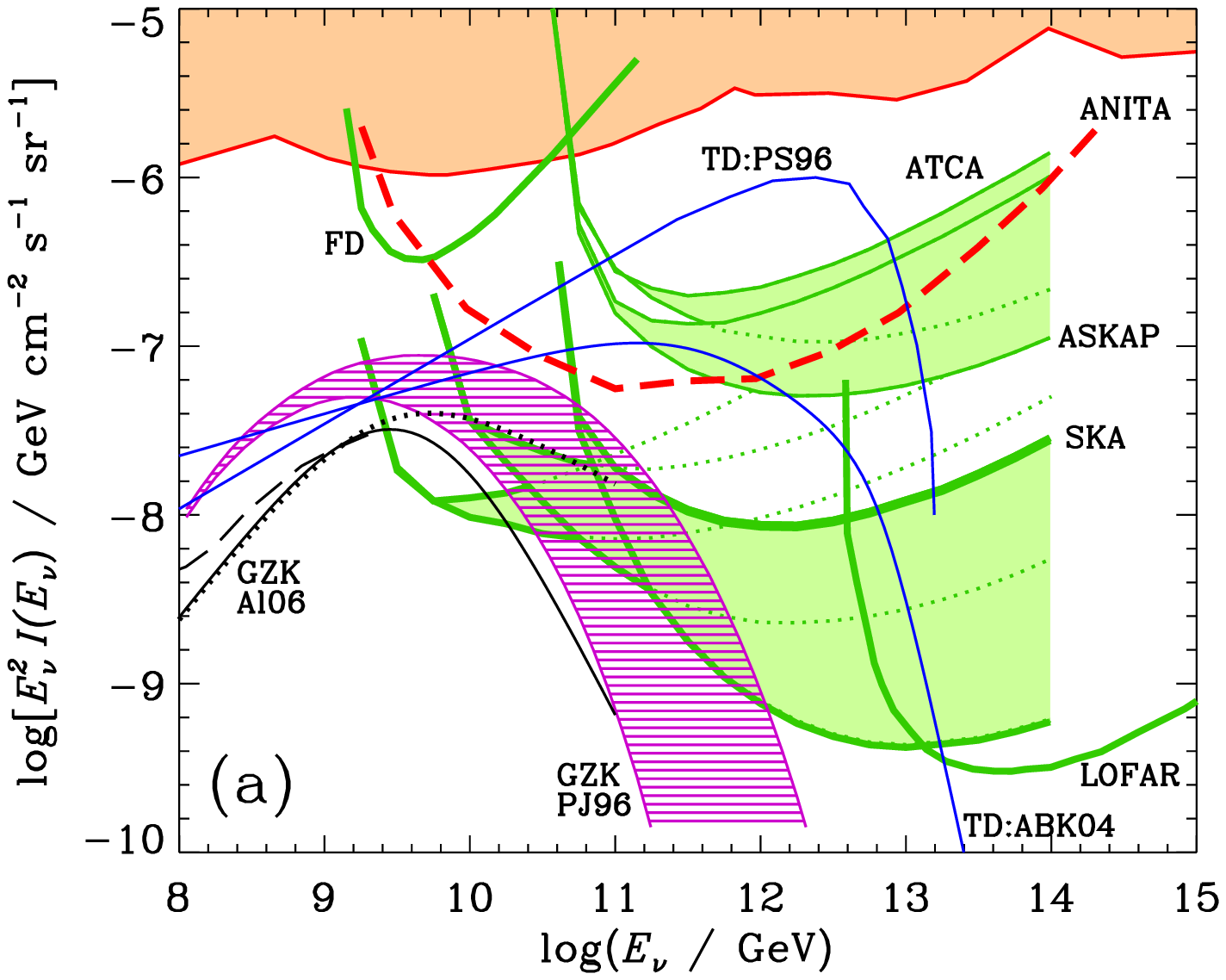, width=95mm}\hspace*{-3em}\epsfig{file=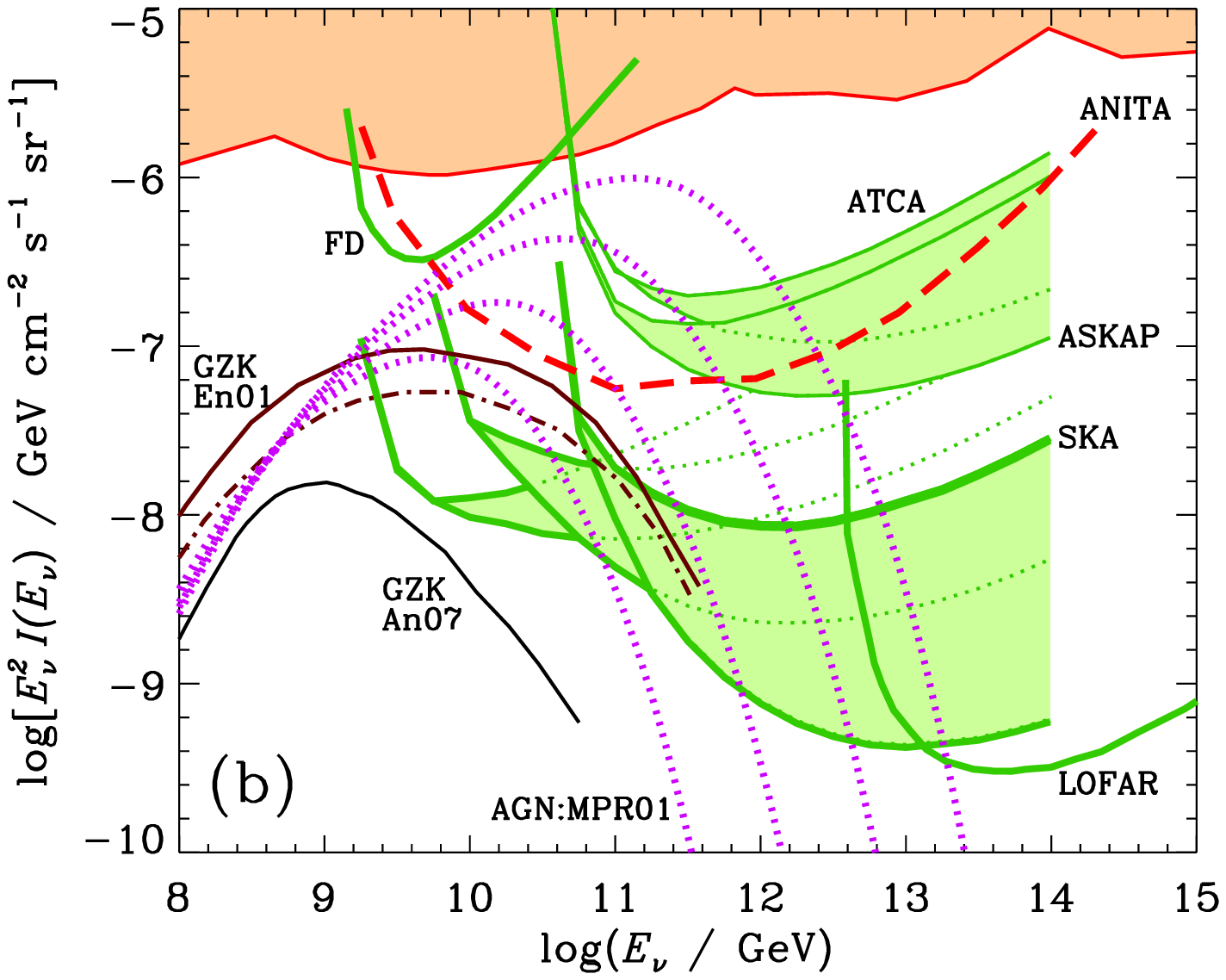, width=95mm}}
\caption{(a) and (b).  Models of UHE neutrino production from GZK
interaction: PJ96--Protheroe \& Johnson \cite{ProtheroeJohnson96}
as expected for UHE CR acceleration in Fanaroff-Riley II radio
galaxies \cite{RachenBiermann} with protons accelerated with
$E^{-2}$ spectrum to $3\times 10^{11}$ and $3\times 10^{12}$~GeV;
En01--Engel et al.  $E^{-2}$ spectrum to $3\times 10^{12}$
(dashed-- EdS cosmology; solid-- $\Lambda$CDM cosmology)
\cite{Engel01}; Al06--Allard et al. \cite{Allard06} with protons
accelerated with $E^{-2.4}$ spectrum to $3\times 10^{11}$~GeV
(dashed), mixed composition with $E^{-2.1}$ to $Z\times 3\times
10^{11}$~GeV (solid), and mixed composition with $E^{-2.1}$
spectrum to $Z\times 3\times 10^{12}$~GeV (dotted); An07--
Anchordoqui et al. \cite{Anchordoqui07} with protons accelerated
with $E^{-2.2}$ spectrum to $4\times10^{11}$~GeV.  Topological
defects (TD) by: PS96--Protheroe \& Stanev
\cite{ProtheroeStanev96}, ABK04--Aloisio, Berezinsky \&
Kachelrei{\ss} \cite{AloisioBerezinskyKachelreiss04} (see
\cite{Berezinsky05}).  Optically thin AGN photoptoduction sources
MPR01 with protons accelerated to $3\times 10^{11}$, $10^{12}$,
$3\times 10^{12}$ and $10^{13}$ GeV, based on Fig.~2b of
Mannheim, Protheroe \& Rachen \cite{MannheimProtheroeRachen01}.
The limit on the total flux of UHE neutrinos (adjusted for all
neutrino flavours) is plotted as the region excluded by past
experiments (shaded area at top).  The projected ANITA limit from their
2006 experiment \cite{Barwick06} (adjusted for balloon flight
duration), and predicted sensitivity for one calendar year of
operation of future experiments to a flux of UHE neutrinos
(adjusted for all neutrino flavours): `FD' Auger Fluorescence
Detectors \cite{Miele05}; LOFAR \cite{Scholten06}; shaded bands
(upper boundary -- limit for 10~m regolith; lower boundary --
10~m regolith plus 2~km sub-regolith) from present work are shown
for ATCA, ASKAP, SKA (left -- dishes, middle -- mid-frequency AA,
right -- low-frequency AA).  The sensitivity for the
mid-frequency AA is only shown as a shaded band where it is lower
than the low-frequency AA shaded band and elsewhere is shown by
thin dotted lines.  Similarly, the sensitivity for the dishes is
only shown as a shaded band where it is lower than the
mid-frequency AA shaded band and elsewhere is shown by thin
dotted lines. }
\label{future_limits}
\end{figure*}

In Fig.\ \ref{future_limits}, we plot the region excluded by past
experiments (shaded area) together with a range of predicted UHE
neutrino fluxes which should be detectable by future experiments.
First we consider models of GZK neutrinos.  Being the only
guaranteed source of UHE neutrinos we shall give a fairly
detailed summary of flux predictions.  The assumptions in all
cases are that the UHE CR are extragalactic, and accelerated with
a power-law spectrum $E^{-\alpha}$ up to some maximum energy
$E_{\rm max}$, often with an exponential cut-off (see
\cite{Protheroe04} for a discussion of cut-offs and pile-ups in
spectra of accelerated particles).  Furthermore the cosmic ray power is
assumed to evolve with redshift $z$, usually in a similar way to
the star formation rate or quasar luminosity function.  Cosmic
rays are propagated through the CMBR by Monte Carlo or other
means taking account of the $z$-dependence of the CMBR and other
target fields, and their flux at $z=0$ is obtained by weighting
with the cosmic ray power evolution model, and integrating of
redshift taking account of cosmological expansion.  The resulting
UHE CR flux is normalized to the observed spectrum, and the same
normalization factor is applied to the GZK neutrino flux
resulting from the same calculation.  In the case of a mixed
composition, i.e.\ protons plus heavy nuclei, it is reasonable to
assume that if protons are accelerated to $E_{\rm max}$, then
nuclei are accelerated to $ZE_{\rm max}$ where $Z$ is the atomic
number.  Generally, the trend is that the higher the maximum
energy and the flatter the spectrum on acceleration, the higher
the GZK neutrino flux.  Also, models with strong evolution of
cosmic ray source power with redshift tend to give a higher GZK
neutrino flux.

One such calculation was made by Protheroe \& Johnson
\cite{ProtheroeJohnson96} of the GZK neutrino flux expected for
the case of UHE CR acceleration in Fanaroff-Riley II radio
galaxies \cite{RachenBiermann} with protons accelerated with an
$E^{-2}$ spectrum to $3\times 10^{20}$ or $3\times 10^{21}$~eV
(boundaries of shaded area in Fig.~\ref{future_limits}a labelled
PJ96).  More recently, a similar calculation was made by Engel et
al.\ \cite{Engel01}, with an $E^{-2}$ spectrum to $3\times
10^{21}$~eV, cosmic ray source evolution of $(1+z)^4$ to $z$=1.9
and constant for $1.9<z$, and Einstein-de Sitter cosmology. Their
result (chain curve in Fig.~\ref{future_limits}b labelled En01)
was almost indistinguishable from that of Protheroe \& Johnson
for the same cut-off.  They found that using the more recent
$\Lambda$CDM cosmology the flux (solid curve in
Fig.~\ref{future_limits}b labelled En01) would be about a factor
of 2 higher.  Even more recently, Allard et al.\ \cite{Allard06}
have made calculations for a variety of spectra and compositions.
We show curves labelled Al06 in Fig.~\ref{future_limits}a, all of
which are for cosmic ray source evolution $(1+z)^4$ to $z$=1 and
constant for $1<z<6$, (i) solid curve -- a mixed composition
accelerated with an $E^{-2.1}$ spectrum to $Z \times 3\times
10^{20}$~eV, (ii) dotted curve -- a mixed composition accelerated
with an $E^{-2.1}$ spectrum to and $Z \times 3\times 10^{21}$~eV,
and (iii) dashed curve -- protons accelerated with an $E^{-2.4}$
spectrum to $3\times 10^{20}$~eV.  The mixed compositions can
give rise to lower neutrino fluxes because photo-disintegration
dominates for heavy primaries, and the resulting nucleons having
lower energy have fewer pion photoproduction interactions from
which neutrinos result.  Similarly, production spectra steeper
than $E^{-2}$ result in fewer neutrinos.  Hence the neutrino flux
predicted by Allard et al.\ is lower than that of Protheroe \&
Johnson and cuts off at lower energy.  
Anchordoqui et al. \cite{Anchordoqui07} consider a range of
maximum energies, compositions and spectra.  Their GZK neutrino
flux is plotted for protons injected with a $E^{-2.2}$ spectrum
to $4 \times 10^{20}$~eV and cosmic ray source evolution
$(1+z)^3$ to $z$=1.9 and constant for $1.9<z<2.7$ -- their
predicted GZK neutrino flux is significantly lower than all the
other predictions of GZK neutrinos.

In addition to these GZK neutrino fluxes, we show two topological
defect (TD) models, a generic TD model just allowed by gamma-ray
and cosmic ray data in 1996 (Protheroe \& Stanev
\cite{ProtheroeStanev96}), and a more recent TD model for
`necklaces' (Aloisio, Berezinsky \& Kachelrei{\ss}
\cite{AloisioBerezinskyKachelreiss04}, see \cite{Berezinsky05}). These are plotted in
Fig.~\ref{future_limits}.  In addition, we show a range of generic
optically thin AGN photoproduction source models with protons
accelerated to $3\times 10^{20}$, $10^{21}$, $3\times 10^{21}$
and $10^{22}$ eV, interpolation based on Fig.~2b of Mannheim,
Protheroe \& Rachen \cite{MannheimProtheroeRachen01}.  While
those models with higher maximum energies may already be ruled
out (on various grounds), they are included to provide some
benchmarks for estimating neutrino event rates.

\subsection{Sensitivity of Future Experiments}

For future experiments, we plot the sensitivity (cm$^{-2}$
s$^{-1}$ sr$^{-1}$ GeV$^{-1}$) to the total neutrino flux over
all flavours, assuming complete flavour mixing. That is, for
experiments sensitive to one flavour only, we multiply the
sensitivity by 3, and for those presenting projected limits, we
remove the statistical uncertainty factor $s_{\rm up}$ by
dividing by 2.3.  To avoid the somewhat arbitrary choice of
`observation time' $t_{\rm obs}$ for future experiments, we use
our best estimates of the mean observation time available in one
calendar year. For lunar Cherenkov observations, this corresponds
to the time the Moon appears above the telescope horizon. Using a
$30^{\circ}$ elevation angle for the horizon of aperture arrays
and $10^{\circ}$ for other radio instruments gives a mean on-time
fraction $\epsilon_{\rm on}$ of $28.8$\% ($105$ days per year)
and $42.6$\% ($156$ days per year) respectively for an instrument
at latitude $\sim$26.5$^{\circ}$ S (applicable to ASKAP and, the
authors' personal bias hopes, the SKA), and slightly less for
ATCA at $30^{\circ}$ S. For LOFAR \cite{Scholten06}, we scale the
expected limit for $30$ days to $17$\% of a year, reflecting the
high latitude (at $\sim$52.5$^{\circ}$ N) and approximate
$30^{\circ}$ horizon, and convert to sensitivity.  For the Pierre
Auger fluorescence detectors (FD), with $\epsilon_{\rm on}=10$\%,
Miele et al.~\cite{Miele05} recently estimated the total
effective aperture to tau leptons.  For the Auger surface
detectors (SD) \cite{auger_nu}, which operate continuously,
$\epsilon_{\rm on}=100$\%.

The resulting sensitivities using $t_{\rm obs}=\epsilon_{\rm on}
\times~1~{\rm year}$ on an UHE neutrino flux for these
experiments are plotted together with predictions of an UHE
neutrino flux in Fig.\ \ref{future_limits}.  Note that our
sensitivity estimates for the SKA are less than previous
estimates submitted to `Square Kilometre Array Design Studies'
\cite{skads}. This is due to a combination of changed assumptions
about the SKA's sensitivity (particularly for the low-frequency
AA), the use of three independent technologies to cover the
critical frequency range (nominally $100$~MHz to $3$~GHz) of
interest to lunar Cherenkov observations, and a different assumed
observation time.

\subsection{Expected Event Rates}

In Table \ref{table4} we give the number of events
expected per calendar year for the UHE neutrino flux models in
Fig.\ \ref{future_limits} and the future lunar Cherenkov
experiments we have simulated.  The first of these possible
experiments to become available will be with ATCA, with the
completion of the CABB (Compact Array Broadband) upgrade in
2009. Since the dishes are relatively large (22~m), the antenna
beam pattern cannot cover the Moon uniformly at high frequencies,
though we find that a limb-pointing configuration (utilising the
full bandwidth of $1$-$3$~GHz) is optimal only at energies
fractionally above the minimum detectable. Since the applicable
energy range is so small, and the effective aperture in this
range less than $0.1$~km$^2$, we exclude a limb-pointing
configuration from our analysis.  In centre-pointing
configuration, utilising the full bandwidth still yields the
greatest aperture near threshold, though above $10^{20}$~eV, the
optimal peak frequency $f_{\rm max}$ is below $2$~GHz. Though
this experiment actually offers a lower collecting area than the
Parkes experiment, this is more than compensated for by the large
bandwidth and ability to simultaneously cover the entire visible
lunar surface near $1$~GHz, and we find an improvement of
approximately an order of magnitude in both threshold and
aperture is expected.

ASKAP, due to be completed in approximately 2011, offers up to
twice the aperture of ATCA --- due to a lower observation
frequency --- but with a similar threshold. The relative utility
of each instrument as an UHE neutrino detector will probably
therefore be determined by other considerations, such as ability
to utilise existing processing power for de-dispersion (likely
favouring ASKAP), usefulness as a platform to develop
technologies scalable to the SKA (also likely favouring ASKAP),
and competition for observation time (favouring ATCA).  Either of
these two instruments will be able to improve on existing limits
from RICE and ANITA-lite in the $10^{20}$--$10^{23}$~eV range in
approximately two calendar months. While the expected number of
events per calendar year is negligible for most of the production
models (see Table \ref{table4}), such observations
would be sensitive to TD models considered by Protheroe \& Stanev
\cite{ProtheroeStanev96}, and the larger flux estimates allowed
under the generic optically thin AGN photoproduction models of
Mannheim, Protheroe \& Rachen \cite{MannheimProtheroeRachen01}.

As expected, the SKA offers a further leap forward in both
threshold and sensitivity beyond the capabilities of ATCA or
ASKAP. The three technology bands (dishes, low- and
high-frequency AAs) are complementary, with the highest aperture
from the low-frequency AA above $10^{20}$~eV, and from the dishes
below $3 \times 10^{19}$~eV. The sensitivity of the
high-frequency AA band ($0.2$-$1$~GHz) is somewhat reduced since
the contribution to total system temperature from lunar thermal
emission incident on a $60$~m diameter cluster of AA tiles is
significant. To a lesser extent this is true for the dishes, but
in the low AA band (70--200 MHz), the Moon will actually appear
colder than the sky due to rising galactic noise.

An apparent contradiction is that the estimated limits from LOFAR
of Scholten et al.\ \cite{Scholten06} are in fact stronger above $10^{22}$~eV by a 
factor of $2$ than those for the SKA low-frequency AA, which will
have a greater collecting area over the frequency range, and thus
\textit{must} be able to set the stronger limit. We emphasise that
this factor is explained by differences in the modelling --- we have
run our simulation using the authors' reported techniques, and found
agreement within the limits of our ability to reproduce their methods.
The primary reason for our more pessimistic result is that
the regolith substrate used here is a less efficient producer and
transmitter of coherent Cherenkov radiation than the uniform regolith
of Scholten et al., while numerous other differences in simulation
techniques, e.g.\ our inclusion of surface roughness, are secondary effects.
We would therefore argue that the LOFAR curve should sit entirely above
that presented here for the SKA low-frequency AA.

The sensitivity provided by the SKA over all technology bands
will allow most predictions of the GZK flux of UHE neutrinos to
be probed in a single calendar year, with the sole exception
being the model of Anchordoqui et al.\ \cite{Anchordoqui07}. The
remaining GZK models will need as little as 1~month (Protheroe \&
Johnson \cite{ProtheroeJohnson96}, $E_{\rm max} = 3 \times
10^{21}$ eV) or as many as $18$ months (Engel \cite{Engel01},
EdS) to be detected / ruled-out at $90$\%
confidence. Importantly, the flux of UHE neutrinos at energies
detectable in the sub-regolith is predicted to be very low in all
GZK models, with such events contributing at most $10$\% of the
simulated detections, so that the detectability of the GZK flux
is insensitive to the nature of the sub-regolith layer. For
models (particularly TD models) predicting a large flux of
neutrinos at the highest energies, the nature of the sub-regolith
becomes very significant, and event rates could be as large as
ten per day. The huge variation in the event rates --- over four
orders of magnitude --- reflects both the current uncertainty in
UHE CR origin, and the size of the parameter space which the SKA
will be able to explore.

Of particular interest is the relative importance of the three
SKA technology bands, with the band exhibiting the highest
individual event rate being dependent on the flux models. Models
where the low-frequency AA will have the highest event rate
predict large fluxes of neutrino signals in all frequency bands,
and will also be (at least marginally) detectable by both ATCA
and ASKAP, suggesting observations below $200$~MHz, purely in
terms of model detection and/or elimination, will be less
critical for UHE neutrino physics. Individually, only the dishes
are likely to detect fluxes consistent with all predictions, bar
that of Anchordoqui et al., in a calendar year.  However, given
the myriad other frequency-dependent issues associated with lunar
Cherenkov observations which we have so-far ignored (see the
discussion), we feel that at this stage it would be unwise to
presume conclusions can be drawn as to which frequency band will
be most important for UHE neutrino observations.

The energy range at which the SKA could set a dominant limit is
almost identical to that of ANITA, reflecting the overlap in
observation frequencies (ANITA observes between $200$-$1200$~MHz)
and the similar geometry of the two experiments. At energies
below $3 \times 10^{18}$~eV (the approximate SKA detection
threshold), the expected flux is high enough to be detected by
both Auger and fixed-volume Antarctic experiments such as
IceCube. So far, our analysis has been purely in terms of
sensitivity to an isotropic flux of UHE neutrinos -- we will
delay a discussion of the arrival direction sensitivity to a subsequent paper.

\section{UHE Cosmic Ray detection with the SKA}
\label{cosmicrays}

In calculating the effective apertures of lunar Cherenkov
experiments to UHE CRs, we found the results for the two methods
of generating surface slopes to diverge for frequencies above a
few hundred MHz, making it impossible to obtain reliable
estimates for ATCA, ASKAP, and the SKA dishes, while for the SKA
high-frequency AA, the two methods gave values different by more
than a factor of two below $10^{21}$~eV. We therefore only
present results for the SKA AAs, and await developments in lunar
surface roughness theory to determine if the sensitivity to UHE
CR at higher frequencies is significant.

\begin{figure*}[htb]
\centerline{\epsfig{file=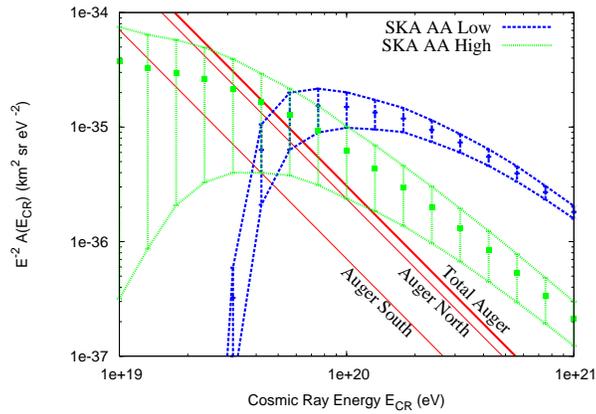,width=8cm}}
\caption{Comparison of the weighted effective aperture (effective
aperture multiplied by fractional on-time) to cosmic rays of the SKA Aperture
Arrays (high and low) with that of a potential 2020 Pierre Auger
project (`Total Auger'), assuming a Northern site (`Auger North')
of $10,000$~km$^2$ in addition to the current observatory (`Auger
South'). The upper and lower bounds come from calculations
respectively assuming independent and unfavourable surface
slopes.}
\label{cr_apps}
\end{figure*}

The conventional experiment currently with the largest UHE CR
aperture is that of the Pierre Auger Observatory. Consisting of a
single $\sim$3000~km$^2$ site in Argentina, by 2020 (when the
full SKA comes on-line), a second, perhaps larger site in the USA
will probably have been completed, with up to $10,000$~km$^2$ of
area. Here we assume sensitivity to all events of zenith angle
$<60^{\circ}$, and $100$\% detection efficiency, giving the total
aperture of both sites to be approximately
$30,000$~km$^2$~sr. For an even comparison, we again weight the
aperture by the fractional on-time $\epsilon_{\rm on}$ ($=1$ for
Auger, and $0.288$ for the SKA AAs). The resulting weighted
effective apertures are plotted in Fig.\ \ref{cr_apps}.

Even using the lower bound for the SKA apertures, the SKA
low-frequency AA could expect a higher CR event rate above
approximately $60$~EeV than the combined Auger
observatories. Coincidentally, this is approximately the energy
at which the most significant anisotropies in arrival directions
have been observed by Auger \cite{AugerScience07}. Since the
aperture increases rapidly with cosmic ray energy, the event rate
will fall much more slowly than with a fixed-aperture experiment
such as Auger. This will allow a greater proportion of CR at the
very highest energies (with the lowest deflection by magnetic
fields) to be observed, although consequently the rate will be
more sensitive to the spectral index at the highest
energies. Measurements of the UHE CR spectrum from Auger
\cite{Yamamoto07, Roth07} indicate a spectrum of $dN/dE \propto
E^{-4.14 \pm 0.42}$ above $10^{19.6}$~eV, giving an annual event
rate for the SKA low-frequency AA above $56$~EeV of between 260
(unfavourable surface slopes, spectral index $-4.56$) and 1050
(independent slopes, spectral index $-3.72$), with the
contributions to the uncertainty from the surface slopes model
and UHE CR spectrum approximately equal.  For comparison, the
rate for a $13,000$~km$^2$ total collecting area (i.e.\ including
the future Northern Hemisphere array) Pierre Auger Observatory is
between about $40$ and $60$ per calendar year (depending only on
spectral index). For any given assumption of surface roughness
and spectral index, the SKA low-frequency AA would detect at
least $\sim 30$ times as many cosmic rays above 56~EeV as the
current Southern Pierre Auger Observatory.

The uncertainty in the aperture of the SKA high-frequency AA is
somewhat greater than that for the low-frequency AA. While the
sensitivity surpasses that of the low-frequency AA mostly only at
energies where the Auger aperture dominates, there is a regime
near 40-60~EeV where the apertures are comparable. In this
regime, the optimal observation band for the low-frequency AA
covers the full instrumental range of $70$--$200$~MHz, while that of
the high-frequency AA is only $200$--$300$~MHz, suggesting that
there could be a significant advantage in simultaneously observing
with both instruments, or pushing up the maximum frequency of the
low-frequency AA to (say) 300 MHz.

It must be stated that we are not arguing the SKA in particular,
or the lunar Cherenkov technique in general, as a replacement to
ground-based UHE CR detectors. Even at the highest energies, the
technique will not provide any compositional measure of the UHE
CR flux, and the energy resolution will probably be poor compared
to current methods. The attraction lies in its ability to gather
unprecedented statistics on the arrival directions of the highest
energy cosmic rays, enabling more accurate statistics in
correlation studies with potential source distributions. We
qualitatively describe methods for determining the arrival
direction, and for distinguishing cosmic rays from neutrinos,
elsewhere --- see James et al.\ \cite{James_icrc07}.  A shift
towards using higher energy CR for correlation studies will
increase the advantage of the SKA with respect to current
detection techniques.

\section{Discussion}
\label{discussion}

There are two broad bases upon which the accuracy of our
estimates can be questioned, the first being our ability to detect
the natural (or `Luna-given') rate of lunar Cherenkov signals
from UHE particle interactions in the Moon, and the second being
our calculations of the signals themselves.

Our results show the potential apertures and event rates for
future radio instruments, corresponding to the limit of thermal
noise. There are many technical hurdles to be overcome to achieve
this, as discussed by McFadden et al.\
\cite{McFadden07}. Limitations on beam-forming capacity impose a
frequency-baseline constraint on real-time triggering if the
entire Moon is to be observed, meaning that the full sensitivity
could only be recovered by setting a high trigger rate and
writing buffered data from long baselines upon triggering.
Coherent de-dispersion of the signal (which gets dispersed by the
Earth's ionosphere, greatly reducing the peak strength) must be
performed in real time prior to triggering, and this requires
both specialised hardware and a very accurate knowledge of the
ionosphere. The baseline limitation is most restricting at high
frequencies, the de-dispersion at low frequencies. Perhaps the greatest
restriction will be obtaining significant observation time to
perform UHE particle physics on a radio instrument. In this
regard, the SKA AAs are least restrictive, since there is the
possibility of forming multiple independent beams and carrying
out lunar Cherenkov observations simultaneously with conventional
radio astronomy by other telescope users.

There are three dominant sources of uncertainty in our
calculations of radio-signals from UHE particle interactions in
the outer layers of the Moon. The UHE neutrino-nucleon
cross-section is very poorly constrained at such high energies
even within the bounds of standard particle physics. To first
order, the effective experimental aperture scales linearly with
this cross-section, and a reduced cross-section may render some
production models undetectable even with the SKA. Also, it is
unlikely lunar Cherenkov experiments alone will break the
degeneracy between flux normalisation and cross-section except
perhaps with a very large number of detections.
In combination with other experiments however, the prospects are
promising, and we view determining the unknown cross-section as a
scientific goal, rather than a theoretical limitation.

Our model of the regolith depth and sub-regolith layer is
relatively poorly constrained by current observations, and could
change with data from the next generation of lunar orbiters. Our
calculated apertures from low-frequency (i.e.\ AA-low and AA-high) observations
of UHE neutrinos only are sensitive to the existence or otherwise
of such a layer, as radiation at high frequencies cannot escape
from great depth, and cosmic rays interact near the surface. The
(high-frequency) SKA dishes dominate the expected event rates
given in Table \ref{table4} for most models of a UHE neutrino flux,
and the exceptions are those where we expect a
high event rate from all detectors from all modelled experiments.
Therefore the question of regolith depth and sub-regolith structure is of less importance.

The main area in which the theory is under-developed is that of
lunar surface roughness. Even at low frequencies, our method to
put bounds on the aperture to UHE CR still allows a factor of
three uncertainty in the case of the high frequency AA at 50~EeV, and prevented accurate modelling at
higher frequencies. Also, our lower bound is not a physically
rigorous bound in the strictest sense of the term. Obscuration of outgoing radiation by
large-scale surface features (`self-shadowing') --- a closely related
problem --- should also be included. Small-scale surface
roughness, which could affect the coherence of radiation over the
shower length, especially at high frequencies where the Moon is
rougher, has so far been ignored, or treated as large-scale
roughness. We hope to address these issues in a future paper.

We have shown that the SKA could utilise the lunar Cherenkov
technique to detect the UHE neutrino flux above $3 \times
10^{18}$~eV under a wide range of production models, and provide
unprecedented statistics on the flux of $>50$~EeV cosmic
rays. Both ATCA and ASKAP could detect or eliminate the most
optimistic UHE neutrino production models in a reasonable
observation time, though we cannot estimate their utility as
cosmic ray detectors.

While the lunar Cherenkov technique alone can not perform
all the science associated with UHE neutrino or CR observations, the
next generation of radio-instruments will in the near-future be
able to make significant contributions to each.

\section*{Acknowledgments}
We wish to thank J.~Alvarez-Mu\~{n}iz for his advice on the
production of coherent Cherenkov radiation in different media,
and also R.~D.~Ekers and R.~A.~McFadden for their advice on the
challenges of using radio instruments for nanosecond pulse
detection. This research was supported under the Australian
Research Council's Discovery Project funding scheme (project
number DP0559991).

\newpage

\begin{table*}
\begin{center}
\begin{tabular}{|  l | c c c c c  |}
\hline
\hline
$r$ (km)    & 0--500    & 500--1000   &  1000--1670   &  1670--1749.99 &  1749.99--1750.00 \\
\hline
\hline
$\rho$ (g cm$^{-3}$) & 8.11 & 3.81 &  3.40 &   3.00    &  1.80 \\
\hline
\end{tabular}
\end{center}
\caption{The values of lunar density, scaled to give the correct lunar mass \cite{Williams04}.}
\label{table1}
\end{table*}


\begin{table}
\begin{center}
\begin{tabular}{|l |c c c c c c c c c |}
\hline
\hline
Medium		& $n$ & $\ell$ & $\theta_C$ & $ \rho$ & $V_0$	& $f_0$ 	& $ C_{\rm EM}$	 & $C_H$	& $E_{\rm LPM}$\\
\hline
\hline
Regolith	& 1.73 & $60 \lambda $ & $54.7^{\circ}$ & 1.8 & $0.0845$ & 2.32	& $4.57^{\circ}$ & $2.40^{\circ}$	& $7.70\times10^{-4}$ \\
Sub-Regolith	& 2.5 & $29 \lambda$ & $66.4^{\circ}$ & 3.0 & $0.0569$ & 2.38	& $4.69^{\circ}$ & $2.46^{\circ}$	& $4.64\times10^{-4}$ \\
\hline
\end{tabular}
\end{center}
\caption{Shower and radio Cherenkov parameters of the regolith and sub-regolith used in the present simulations. Units: $ \rho$ (g/cm$^3$), $V_0$ (V/MHz), $f_0$ (GHz), $E_{\rm LPM}$ (EeV).}
\label{table2}
\end{table} 


\begin{table*}
\begin{center}
\begin{tabular}{| c l | c c c c c|}
\hline
\hline
Instrument	& 		& $D$ (m) & $N$ & $S_0$  (m$^2/$K) & $f_{\rm min}$ (GHz) & $f_{\rm max}$ (GHz)\\
\hline
\hline
Parkes	&	Hi		& 64 	& 1 & 69 & 1.475 & 1.575\\
	&	Lo		&  	&   &      & 1.275 & 1.375\\
GLUE	& DSS14	LCP		& 70	& 1 & 82.5 & 2.18 & 2.22  \\
	&  DSS14 RCP Hi		& 	&   & 	   & 2.2 & 2.275 \\
	&  DSS14 RCP Lo		& 	&   &      & 2.125 & 2.2 \\
	& DSS13	Hi		& 34	& 1 & 20.5 & 2.2 & 2.275 \\
	&  DSS13 Lo		& 	&   &      & 2.125 & 2.2 \\
Kalyazin 	&		& 64 	& 1 & 69 & 2.25 & 2.35 \\
ATCA		&		& 22	& 6 & 61 & 1 & 3 \\
ASKAP 		&		& 12	& 30 & 77.6 & 0.7 & 1.8\\
SKA & AA Low 			& $60$ & 154 & 4000 & 0.07 & 0.2 \\
 & AA High 			& $60$ & 154 & 10000 & 0.2 & 1 \\
 & Dishes 			& 15 & 2476 & 10000 & 0.7& 3 \\
\hline
\end{tabular}
\end{center}
\caption{Parameters of radio instruments used in the modelling of both
past and future lunar Cherenkov experiments. Respectively, these parameters
are dish diameter / AA cluster size $D$ (m), number of AA tile clusters or
dishes $N$, base sensitivity $S_0$ ($=A_{\rm eff}/T_{\rm sys}$)
before accounting for lunar emission, and frequency range of the
instruments (triggering frequencies will in general be more
restricted for the future experiments). The frequency range is also the maximum bandwidth, except for ASKAP
which will have a maximum bandwidth of $300$~MHz. For GLUE, when in defocused mode, DSS14 was treated as DSS13 in
terms of $D$ and $S_0$.}
\label{table3}
\end{table*}

\newpage

\begin{table*}
\begin{center}\begin{small}
\begin{tabular}{| c c | c c c c c c|}
\hline
Author & Model & ATCA & ASKAP  & AA-low & AA-high & Dishes & Full SKA\\
\hline
\hline
\multirow{2}{*}{PS96} 	& 					&	20 & 40 & 3140 & 950 & 290 & 3200\\
&								&	10 & 14 & 240 & 108 & 73 & 266\\
\hline
\multirow{2}{*}{ABK04} 	&					&	1.6 & 3.3 & 180 & 90 & 49 & 206\\
&								&	1.0 & 1.5 & 21 & 16 & 19 & 34\\
\hline
\hline
\multirow{4}{*}{PJ96} & \multirow{2}{*}{$3\times 10^{20}$ eV}	&	0 & 0 & 0.1 & 1.5 & 6.3 & 6.4 \\
	& 							&	0 & 0 & 0.06 & 1.0 & 5.2 & 5.2 \\
\cline{3-8}& \multirow{2}{*}{$3\times 10^{21}$ eV}		&	0.18 & 0.33 & 9 & 18 & 29 & 35\\
&								&	0.16 & 0.23 & 2.6 & 7 & 18 & 19\\
\hline
\hline
\multirow{4}{*}{En01} &\multirow{2}{*}{EdS}			&	0 & 0  & 0.03 & 0.3 & 1.7 & 1.7\\
	&							&	0 & 0  & 0.01 & 0.2 & 1.5 & 1.5\\
\cline{3-8}&\multirow{2}{*}{$\Lambda$CDM} 			&	0 & 0  & 0.04 & 0.5 & 2.9 & 2.9\\
	&							&	0 & 0  & 0.02 & 0.34 & 2.5 & 2.5\\
\hline
\hline
\multirow{4}{*}{Al06} &\multirow{2}{*}{$3\times 10^{20}$ eV} 	&	0 & 0 & 0.1 & 0.8 & 3.3 & 3.4\\
	&							&	0 & 0 & 0.04 & 0.5 & 2.7 & 2.7\\
\cline{3-8}&\multirow{2}{*}{$3\times 10^{21}$ eV}		&	0.18 & 0.35 & 21 & 12 & 11 & 30\\
 &								&	0.11 & 0.17 & 2.3 & 3.0 & 6.4 & 7.9\\
\hline
\hline
\multirow{2}{*}{An07}				 	&	&	0 & 0  & 0 & 0.02 & 0.24 & 0.24\\
							&	&	0 & 0  & 0 & 0.02 & 0.22 & 0.22\\
\hline
\hline
\multirow{8}{*}{MRP}&\multirow{2}{*}{$3\times 10^{20}$ eV}	&	0.02 & 0.03 & 0.6 & 5.7 & 18 & 18\\
 &								&	0.02 & 0.02 & 0.3 & 3.4 & 14 & 14\\
\cline{3-8} &  \multirow{2}{*}{$10^{21}$ eV}			&	0.29 & 0.52 & 12 & 36 & 59 & 66\\
&								&	0.27 & 0.4 & 4.4 & 16 & 37 & 38\\
\cline{3-8} &\multirow{2}{*}{$3\times 10^{21}$ eV}		&	2.6 & 5.0 & 156 & 194 & 182 & 293\\
&								&	2.2 & 3.3 & 38 & 58 & 88 & 105\\
\cline{3-8} &\multirow{2}{*}{$10^{22}$ eV}			&	14 & 28 & 1220 & 820 & 480 & 1480\\
&								&	10 & 15 & 190 & 170 & 180 & 300\\
\hline
\end{tabular}
\end{small}
\end{center}
\caption{Expected number of neutrino events per calendar year from
lunar Cherenkov experiments for the different models of the UHE
neutrino flux plotted in Fig.\ \ref{future_limits}, including
(upper value) and excluding (lower value) the sub-regolith. A `0'
implies an expected annual event rate of less than $0.05$. The
three SKA technology bands have been estimated separately, with
the total calculated using the highest aperture at a given
energy.}
\label{table4}
\end{table*}

\end{document}